\documentclass[amsmath,amssymb,epsfig,aps,prx,twocolumn]{revtex4-1}
\usepackage{amsmath}
\usepackage{float}
\usepackage{epsfig}
\usepackage{graphicx}
\usepackage{color}
\usepackage{amssymb}
\usepackage{epstopdf}
\begin{document} 
\title{ Cold-atom-dimer reaction rates with $^4$He, $^{6,7}$Li and $^{23}$Na} 
\author{M. A. Shalchi$^1$, M. T. Yamashita$^{1}$, T. Frederico$^{2}$ and Lauro Tomio$^{1}$\\ }
\affiliation{
$^{1}$ Instituto de F\'\i sica Te\'orica, Universidade Estadual Paulista, 01405-900 S\~{a}o Paulo, Brasil.\\
$^{2}$ Instituto Tecnol\'ogico de Aeron\'autica, DCTA, 12228-900 S\~ao 
Jos\'e dos Campos, Brasil.
}
\date{\today}
\begin{abstract}  
Atom-dimer exchange and dissociation reaction rates are predicted for different combinations of two 
$^4$He atoms and one of the alkaline species among $^{6}$Li, $^{7}$Li and $^{23}$Na, by using three-body scattering formalism
with short-range two-body interactions. Our study was concerned with low-energy reaction rates in which the $s-$, $p-$ 
and $d-$ wave contributions are the relevant ones.  The $^4$He is chosen as one of the atoms in the binary mixture, in view 
of previous available investigations and laboratory accessibilities. Focusing on possible experimental cold-atom realizations with 
two-atomic mixtures,  in which information on atom-dimer reaction rates can be extracted, we predict the
occurrence of a dip in the elastic reaction rate for colliding energies smaller than 20 mK, when the dimer is the
$^4$He$^{23}$Na molecule. We are also anticipating a zero in the elastic $p-$wave contribution for the 
$^4$He + $^4$He$^7$Li and $^4$He + $^4$He$^{23}$Na reaction processes. 
With weakly-bound molecules reacting with atoms at very low colliding energies, we interpret our results on the light of 
Efimov physics which supports model independence and robustness of our predictions. Specific sensitivities on the effective 
range were evidenced, highlighted by the particular inversion role of the $p-$ and $d-$waves in the atom exchange 
and dissociation processes.
\end{abstract}

\maketitle

\section{Introduction}\label{intro}
The fast development of methods to control atom and molecules in ultra-cold experiments, which have followed after
the realization that Feshbach resonance techniques~\cite{1958Feshbach} could be used to manipulate the 
two-body interaction~\cite{1998Inouye,1999Timmermans,2010Chin}, made possible to  emerge ultra-cold chemistry 
as a new field of interest with an intense research activity in recent years. 
On this regard, some previous reviews can be found on collisions with ultra-cold atoms and related discussions on 
possible experimental investigations~\cite{1995Lett,1995Heinzen,1999Bagnato,1999Stwalley}, which were followed 
by Ref.~\cite{2004Stwalley}, where the basis are settled for investigations on collisions and reactions with ultracold molecules. 
This work \cite{2004Stwalley}, in which it was concluded that the magnitude of elastic collision cross sections depends 
more on the mass and symmetry than on the interaction, came just after the experimental realization of Bose-Einstein 
condensation of Cooper-paired molecules of $^6$Li$_2$~\cite{2003Jochim} and $^{40}$K$_2$~\cite{2003Greiner}.
Inelastic molecule-molecule and molecule-atom collisions were also characterized within the study of dissociation and 
decay of ultra-cold sodium molecules driven by the Feshbach mechanism~\cite{2004Mukaiyama}. {The 
investigations on ultra-cold molecules via Feshbach mechanism can be exemplified by recent production of 
$^{23}$Na$^{87}$Rb molecules reported in Ref.~\cite{2016Guo} and the more recent investigation on ultra-cold 
$^{23}$Na$^{39}$K ground-state molecules~\cite{2020Voges}, which can provide conditions for fully controlled studies with
ultra-cold molecular collisions.
Furthermore,  it was recently  reported  
 progress   with extreme mass imbalanced Fermionic mixtures where for the first time was observed magnetic Feshbach resonances 
 in $^6$Li and $^{173}$Yb ultra-cold atoms~\cite{2020Green}.}

With the complex structure of molecules, new opportunities can also be opened for research, such as the realization 
of quantum fluids of bosons with anisotropic interactions, as well as quantum coherent chemistry by considering 
atom-molecule conversions. 

Some pioneer experimental investigations with trapped ultracold atom-molecule collisions have been reported in
2006, in Refs.~\cite{2006Staanum,2006Zahzam}, by considering cesium (Cs) atoms in Cs$_2$ molecules. These works 
are followed by experiments with molecular collisions with tunable halo dimers, exploring four-body processes with 
identical bosons~\cite{2008Ferlaino}, inelastic collision rates of $p-$wave $^6$Li$_2$ molecules~\cite{2008Inada},
RbCs molecule collisions with Rb and Cs atoms~\cite{2008Hudson}.
Applications in cold controlled chemistry, with the possibilities to manipulate molecule 
collisions by electromagnetic fields,  were discussed in Refs.~\cite{2008Krems,2009Carr,2010Ospelkaus}, which 
together with references therein 
are covering both theoretical and experimental aspects in the first stages of investigations along these lines. 
As shown in these works, collisions of molecules with temperatures below 1 Kelvin can be 
manipulated by external electromagnetic fields. In  Ref.~\cite{2010Ospelkaus}, by considering potassium-rubidium 
molecules, it was also pointed out that, even when the cooled molecules have no energy to collide, 
 exothermic atom exchange reactivity processes can occur through quantum mechanical tunneling.
Nowadays, the reported experimental setups are showing an increasing level of control on atomic and molecular states, 
as one can trace from ongoing experimental and theoretical investigations, which can be exemplified by
Refs.~\cite{2008Klempt,2009Bell,2010Ni,2010Knoop,2010Lompe,2010Nakajima,2011Miranda,2012Quemener,2012Park,2013Wang}.
For more recent experimental activities reported in the last five years, on ultracold atom-dimer collisions, by using tunable
Feshbach resonances, we should also mention the
Refs.~\cite{2017Rui,2018Hoffmann,2018Gao,2019Yang,2019Li,2019Wang,2020Reynolds,2020Makrides}.

\begin{figure*}[thb]
\includegraphics[width=18cm]{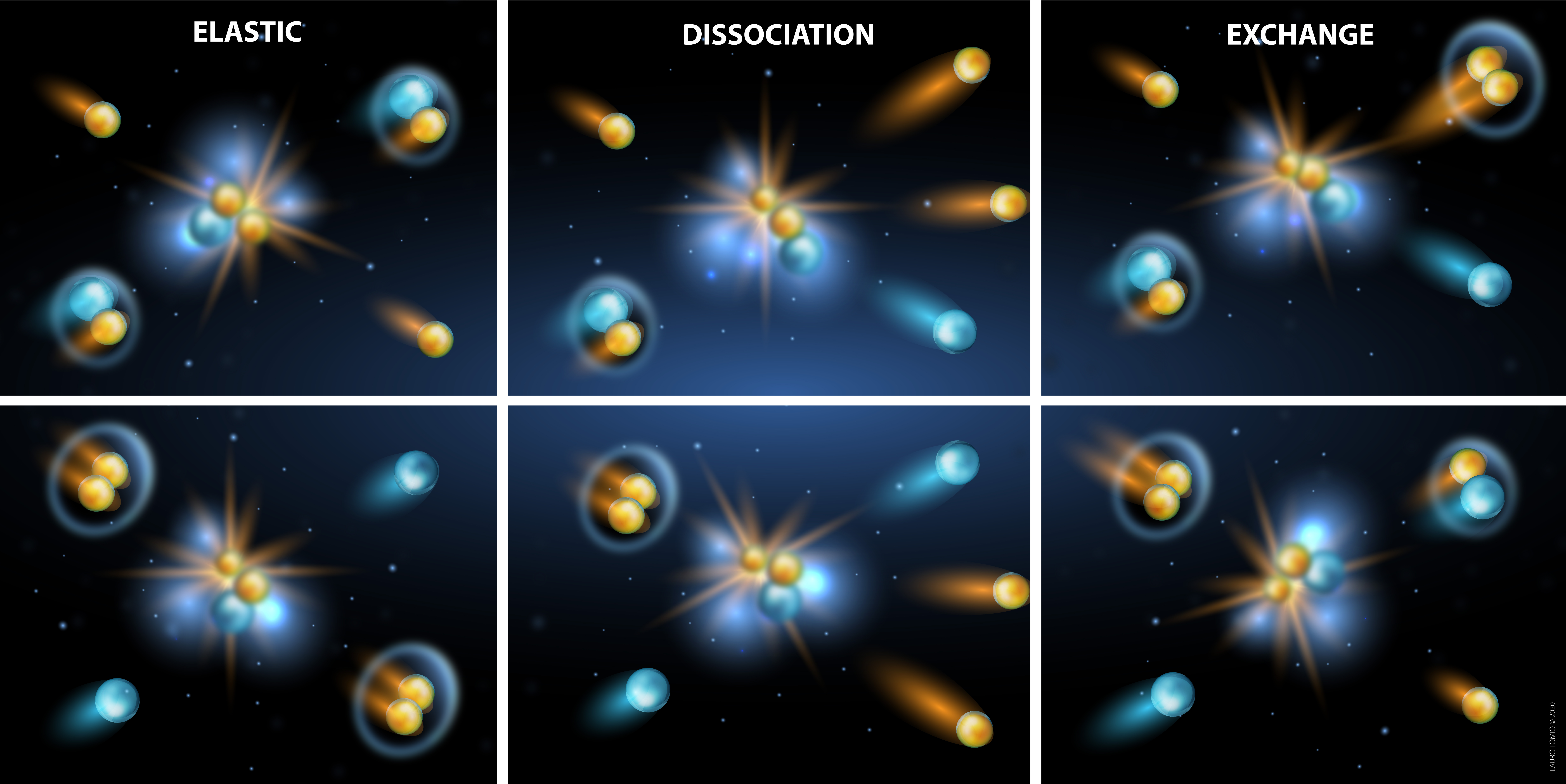}
\vspace{-.3cm}
\caption{\bf Illustration of the atom-dimer reaction (for each panel, the entrance channel is in the left side,   
with the exit channel at the right side). The $^{4}$He is represented by the identical particles, with the 
dimers represented by $^{4}$He$\beta$ ($\beta=$ $^{6}$Li, $^{7}$Li or $^{23}$Na) and $^{4}$He$_2$ bound states. 
Assuming the $^{4}$He$\beta$ is more bound than $^{4}$He$_2$, the atomic exchange reaction
is endothermic in the upper-right panel;  and exothermic in the lower-right one.}
\vspace{-0.5cm}
\label{fig00}
\end{figure*}

Our study is also  motivated by the interest on verifying manifestations of Efimov physics~\cite{Efimov} in atomic 
reactions, following Ref.~\cite{2018ShalchiPRA98}, which could support previous theoretical studies with three-body weakly-bound systems~\cite{1977lim,1978lim,1978Huber,1983lim,1984CG,1998Yuan,1999-Amorim,2014Stipa,2019Stipa}. 
In particular, the existence of a weakly-bound excited state in helium was stablished in Ref.~\cite{1984CG}. 
This matter was further investigated in the Efimov physics context in  Refs.~\cite{1985Huber,1996Esry,1997Kolganova,2001Motovilov,2004Kolganova,2000Delfino,2000Gianturco,2002Gianturco,2012Roudnev,2014Wu,2017Kolganova,2018Kolganova},  
motivated by the remarkable small binding energy of the $\,^{4}$He dimer: $B_{^4\rm{He}_2} =$ 
1.31 mK,  which was first reported independently in Refs.~\cite{1993Luo,1994Schollkopf}, being more recently
confirmed experimentally~\cite{2000-Hedimer,2015-kunitski}. 
The experimental success in verifying such long-time theoretical prediction, together with the results 
of previous experimental investigations of Efimov physics in cold atom laboratories~\cite{2009Zaccanti,2009Pollack,2009Knoop}, 
which are extended to mixed atomic-molecular combinations~\cite{2010Barontini,2013Bloom,2014Pires,2016Ulmanis},  
became highly motivating for more deeper studies with single or mixed atomic 
species~\cite{BringasPRA04,2010wang,2011levinsen,2012Garrido,2014blume,2018Shalchi}.
Quite remarkable are the advances in the laboratory techniques, such that one can even consider the possibility to
alter the two-body interaction by using Feshbach resonance mechanisms (originally proposed in the nuclear physics 
context)~\cite{1999Timmermans}. The possibility of tuning the two-body scattering length in ultracold atomic experiments  
 can alter in an essential way the balance between the non-linear first few terms of the mean-field description,  as it was
explored when modeling the atomic Bose-Einstein condensation~\cite{2000Gammal}.  

The above mentioned experimental possibilities in cold chemistry  laboratories and the interest to find further 
manifestations of the universal Efimov physics,
motivate our study focusing on dissociation and atom exchange reactions with weakly-bound atom-molecular systems. 
Our aim is to explore the associated universal properties, which emerge in the studies of the reaction mechanisms of 
cold atoms with weakly-bound molecules. 
By following a previous investigation on $s-$wave scattering properties with the atomic species 
$^4$He, $^{6}$Li, $^{7}$Li and $^{23}$Na~\cite{2018ShalchiPRA98}, with our present study 
we are providing a detailed analysis on atom-dimer dissociation and exchange reactions,   
in which we discuss and quantity the $s$, $p$ and $d$ lower partial-wave contributions to the associated reaction rates.
The choice of these four atomic samples relies on the fact that the dimer and trimer binding 
energies are enough known, either from available realistic potential model calculations or from experiments as in the 
case of $^4$He$_2$, which  provide support to our theoretical predictions.
In this introduction we are including a pictorial illustration of the reactions that we are considering
 (see Fig.~\ref{fig00}), by assuming the 
common case in which the $^4$He$\beta$ molecule ($\beta=$ $^{6}$Li, $^{7}$Li or $^{23}$Na) is 
more bound than the $^4$He$_2$ dimer. For the entrance channel, in the upper row of this illustration we 
have the collision of a single $^4$He atom with the $^4$He$\beta$ molecule; in the lower row, the single  $\beta$ 
atom collides with the $^4$He$_2$ dimer.

 In the present work, the Faddeev formalism was adopted~\cite{1965Fadd} with renormalized zero-range 
(ZR)~\cite{1997-Amorim} and finite-range  Yamaguchi~\cite{1954Yamaguchi} separable (FR) two-body interactions, 
using as inputs the available results reported in Ref.~\cite{2017Suno} for the dimer and trimer binding energies 
obtained from realistic interactions. These realistic interaction binding energies reported in Ref.~\cite{2017Suno} 
are selected from different potential models discussed in the literature, being guided by the corresponding available 
experimental results.

In the next Sect.~\ref{sec2}, we present the basic three-body formalism for two-atom mixtures, in which our main focus 
is on the case of atom-dimer collision, for separable two-body interactions. Details on reaction rates and existent models 
are presented in Sect.~\ref{sec3}. In the next three sections (\ref{sec4}, \ref{sec5} and \ref{sec6}) we have separated the 
corresponding results for the mixtures of $^4$He with $^6$Li, $^7$Li and $^{23}$Na, respectively.
Our final remarks and conclusions are presented in Sect.~\ref{conclusion}. In addition, three appendices 
are added to detail the three-body framework.  

\section{Three-body framework }\label{sec2}
The standard Faddeev formalism for three-body systems~\cite{1965Fadd,AGS,1983Glockle} with 
two atomic species interacting through one-term separable potentials including the zero-range one is presented in this section,  
by considering the elastic 
and rearrangement scattering amplitudes, from the collision of one of the atoms with  a dimer formed by the remaining ones.
(The detailed derivation of the scattering equations are presented in 
Appendix \ref{app:atom-dimer-eqs}.)
As explained in our motivation for this study (with three-body system having two distinct atoms), we  
choose the identical two particles, labelled $\alpha$, as the $^4$He atom, with the third atom,  
labelled $\beta$,  being $^7$Li,  $^6$Li, or $^{23}$Na. 
This particular choice is related to the available data obtained from experimental and well-known potential models 
for the three-body and two-body binding energies, when assuming different combinations of the corresponding 
three-particle atom-dimer systems.  Our motivation also relies on the actual interest in the production of ultracold 
molecular systems~\cite{2016Guo,2020Voges}, and in the recent investigations on binary molecular collisions~\cite{2017Rui,2018Hoffmann,2018Gao,2019Li,2020Reynolds}, together
with experimental studies with these atomic samples~\cite{2013Wang,2020Makrides}.

In this study, we assume that the selected three-body system $(\alpha\alpha\beta)$ is always bound,  
as well as the possible subsystems $(\alpha\beta)$ and $(\alpha\alpha)$, with the corresponding available 
data being used as inputs coming from different realistic potential models (given by previous available studies),
together with existent experimental data.
So, we assume as fixed the  $^4$He$_2$ binding energy and corresponding scattering length, such that 
$E_{\alpha\alpha}=-B_{\alpha\alpha}=-1.31$mK and $a_{\alpha\alpha}=$100\AA.
For the other input binding energies, we select two specific models which have been discussed in 
Ref.~\cite{2018ShalchiPRA98}, that we found enough representative to explore the sensibility of our results on 
dimer binding energies variation. As it will be discussed, by using different potential models, we expect our 
results on reaction rates, together with possible future experimental data, be useful  in selecting the
appropriate two-body potential model.

\subsection{Bound-state $\alpha\alpha\beta$ three-body system}
We start by recovering some details from Ref.~\cite{2018ShalchiPRA98}, in order to fix the notation of the formalism, 
summarizing the $s-$wave three-body bound-state coupled equation when considering separable potentials with
all sub-systems being bound.
We assume units such that  $\hbar=1$ (with energies given in mK), with $m\equiv m_\alpha = m_{^4{\rm He}}$ and 
a mass ratio defined by $A\equiv m_\beta/m_\alpha$. So, for the reduced masses we have $\mu_{\alpha\alpha}=m/2$ 
and $\mu_{\alpha\beta}=Am/(A+1)$ for the $\alpha\alpha$ and $\alpha\beta$ subsystems, respectively. The corresponding 
three-body reduced masses are given by  $\mu_{\alpha(\alpha\beta)}=m(A+1)/(A+2)$ for the $\alpha- (\alpha\beta)$; and 
$\mu_{\beta(\alpha\alpha)}=m(2A)/(A+2)$ for the $\beta- (\alpha\alpha)$.
The bound-state energies for the two- and three-body systems are given by $E_{\alpha\alpha}\equiv -B_{\alpha\alpha}$, 
$E_{\alpha\beta}\equiv -B_{\alpha\beta}$ and $E_3=-B_3$, respectively; with the energies of the $s-$wave elastic colliding 
particle given by $E_{k_\alpha}\equiv$ ${k_\alpha^2}/{[2\mu_{\alpha(\alpha\beta)}]}\equiv$ $E_3-E_{\alpha\beta}$
and $E_{k_\beta}\equiv$ ${k_\beta^2}/{[2\mu_{\beta(\alpha\alpha)}]}\equiv$ $ E_3-E_{\alpha\alpha}$.
The extension to higher partial waves will be done along the following subsection. 
In this case, the Faddeev coupled equations are reduced to integral equations for the momentum space spectator 
functions of the particles $\alpha$ and $\beta$, which are given by  $\chi_{\alpha}({\bf q};E_3)$ and $\chi_{\beta}({\bf q};E_3)$. 
Considering the $s-$wave case, by redefining these functions, as
$\chi_{j=\alpha,\beta}(q;E_3)\equiv {\overline{\chi}_{j}(q;E_3)}/{(q^2+|k_j^2|)},$
the corresponding coupled formalism is given by
\begin{eqnarray}\label{eq:2}
 \overline{\chi}_{\alpha}(q;E_3)&=&\overline{\tau}_{\alpha}(q;E_3)\int_0^{\infty} dk k^2 \left[K^0_2(q,k;E_3)
 {\overline{\chi}_{\alpha}(k;E_3) \over(k^2+|k_\alpha^2|)}\right. \nonumber \\ 
 &+&\left. K^0_1(q,k;E_3){\overline{\chi}_{\beta}(k;E_3)\over (k^2+|k_\beta^2|)}\right], \\
 \overline{\chi}_{\beta}(q;E_3)&=&\overline{\tau}_{\beta}(q;E_3)\int_0^{\infty}dk k^2 K^0_1(k,q;E_3)
 {\overline{\chi}_{\alpha}(k;E_3)\over (k^2+|k_\alpha^2|)}
\nonumber ,\end{eqnarray}
where $ K^0_1(q,k;E_3)$ and $K^0_2 (q,k;E_3)$ are the appropriate momentum space kernels, which will be 
explicitly given according to the kind of form-factors one considers for the two-body interaction. 
 The respective two-body t-matrix elements for the $\alpha\beta$ and
$\alpha\alpha$ bound subsystems are  $\overline{\tau}_{\alpha}(q;E_3)$ and $\overline{\tau}_{\beta}(q;E_3)$,
in which $\overline{\tau}_j$ and the $s-$wave kernels $K^0_{1,2}$ will be appropriately extended to 
$\ell$ partial waves in  the Appendix \ref{app:kernels}, when considering the specific potential models. 

\subsection{Atom-dimer collision}\label{subsec:atomdimercollision}
 By considering atom-dimer collision with two different atomic species,  
we have three separate channels in the continuum: 
elastic scattering, rearrangement (exchange reaction),  and breakup (dissociation reaction).
 For the scattering of $\alpha$ by the $\alpha\beta$ bound subsystem, by 
 using the symmetry properties applied to identical bosons, the $\ell$-partial-wave scattering amplitudes,
 $h_{\alpha}^\ell$ and $h_{\beta}^\ell$, with one-term separable potentials, can be written as follows 
 (see Appendices \ref{app:atom-dimer-eqs} and \ref{app:kernels}):
 
 {\small
\begin{eqnarray}\label{eq:3a}
&&h^\ell_{\alpha}(q;E_3)=\overline{\tau}_{\alpha}(q;E_3)\Bigg\{\frac{\pi}{2}K^\ell_2(q,k_\alpha;E_3)
+\int_0^{\infty} dk k^2 \\
&&\hspace{-.0cm}\times\Bigg[K^\ell_2(q,k;E_3) {h^\ell_{\alpha}(k;E_3)\over (k^2-k_\alpha^2-{\rm i}\epsilon)}
 +K^\ell_1(q,k;E_3)\frac{h^\ell_{\beta}(k;E_3)}{q^2-k_\beta^2-{\rm i}\epsilon}\Bigg]\Bigg\},\nonumber 
 \end{eqnarray}
 \begin{eqnarray}\label{eq:3b}
 h^\ell_{\beta}(q;E_3)&=&\overline{\tau}_{\beta}(q;E_3)\Bigg\{\frac{\pi}{2}K^\ell_1(k_\alpha,q;E_3)\\
&&+\int_0^{\infty}dk k^2 K^\ell_1(k,q;E_3) {h^\ell_{\alpha}(k;E_3)\over(k^2-k_\alpha^2-{\rm i}\epsilon)}\Bigg\}\, 
\nonumber
,\end{eqnarray}
}where $k_{\alpha}$ and $k_{\beta}$ are the on-shell momenta (already defined), with 
$h_{\alpha}^\ell(k_\alpha;E_3)$ representing the on-shell elastic amplitude and $h_{\beta}^\ell(k_\alpha;E_3)$ the 
corresponding on-shell atom-exchange amplitude. 
$K_{1,2}^\ell$ which have been introduced (for $\ell=0$) in Eq. \eqref{eq:2} are momentum space 
kernels in $\ell$-wave.
From the scattering amplitudes, we can obtain the cross sections with the corresponding reaction rates. 
For the elastic scattering, we obtain (see Appendix  \ref{app:xsection})
\begin{eqnarray}
\sigma_{el}(E_{k_\alpha})=4\pi\sum_\ell{(2\ell+1)|h^\ell_{\alpha}(k_\alpha;E_3)|^2} \, ,
\end{eqnarray}
with the exchange reaction cross section given by
{\small\begin{eqnarray}
\sigma_{ex}(E_{k_\alpha})&=&\frac{2\pi f_{\alpha}^2}{f_{\beta}^2}
\sqrt{\frac{\mu_{\alpha(\alpha\beta)}}{\mu_{\beta(\alpha\alpha)}}
\left(1-\frac{E_{\alpha\alpha}-E_{\alpha\beta}}{E_{k_\alpha}}\right)}
\nonumber \\ &\times &
\sum_\ell{(2\ell+1)|h^\ell_{\beta}(k_\alpha;E_3)|^2} \, ,
\end{eqnarray}
}where the on-shell $h_{\alpha}^\ell$ and $h_{\beta}^\ell$ are obtained from Eqs.~\eqref{eq:3a} and 
\eqref{eq:3b}.

For the scattering of particle $\beta$  by the $\alpha\alpha$ bound subsystem, 
the scattering amplitudes can be calculated by the following coupled equation 
(see  Appendices  \ref{app:atom-dimer-eqs} and \ref{app:kernels}):
{\small
\begin{eqnarray}\label{eq:4a}
&&h^\ell_{\alpha}(q;E_3)=\overline{\tau}_{\alpha}(q;E_3)\Bigg\{\pi K^\ell_1(q,k_\alpha;E_3)
+\int_0^{\infty} dk k^2 \\
&&\hspace{-.0cm}\times\Bigg[K^\ell_2(q,k;E_3) {h^\ell_{\alpha}(k;E_3)\over (k^2-k_\alpha^2-{\rm i}\epsilon)}
 +K^\ell_1(q,k;E_3)\frac{h^\ell_{\beta}(k;E_3)}{q^2-k_\beta^2-{\rm i}\epsilon}\Bigg]\Bigg\},\nonumber\\
&& h^\ell_{\beta}(q;E_3)=\overline{\tau}_{\beta}(q;E_3)\int_0^{\infty}dk k^2 K^\ell_1(k,q;E_3) 
{h^\ell_{\alpha}(k;E_3)\over(k^2-k_\alpha^2-{\rm i}\epsilon)}\label{eq:4b}.\end{eqnarray}
}
Here $h_{\alpha}^\ell$ represents the exchange reaction and $h_{\beta}^\ell$ represents the elastic scattering.
In this case elastic and exchange reaction cross sections can be calculated by (see Appendix  \ref{app:xsection}):
\begin{eqnarray}
\sigma_{el}(E_{k_\beta})=\pi\sum_\ell{(2\ell+1)|h^\ell_{\beta}(k_\beta;E_3)|^2} \, ,
\end{eqnarray}
and 
\begin{eqnarray}
 \sigma_{ex}(E_{k_\beta})&=&\frac{2\pi f_{\beta}^2}{f_{\alpha}^2}\sqrt{\frac{\mu_{\beta(\alpha\alpha)}}
 {\mu_{\alpha(\alpha\beta)}}}\sqrt{1-\frac{E_{\alpha\beta}-E_{\alpha\alpha}}{E_{k_\beta}}}\nonumber \\
 &\times &\sum_\ell{(2\ell+1)|h^\ell_{\alpha}(k_\beta;E_3)|^2} \, ,
\end{eqnarray}
where the on-shell $h_{\alpha}^\ell(k_\beta)$ and $h_{\beta}^\ell(k_\beta)$ are calculated from 
Eqs.~\eqref{eq:4a} and \eqref{eq:4b}.  The $f_{\alpha}$ and $f_{\beta}$ factors, together with detailed derivation 
of the scattering amplitudes and cross sections for each potential, are described in the Appendix~\ref{app:xsection}. 

The scattering amplitude for breakup or dissociation reaction can also be calculated by the summation over half-off-shell 
results of $h_{\alpha}$ and $h_{\beta}$ using the Eqs.~\eqref{eq:3a} and \eqref{eq:3b} (for $\alpha - \alpha\beta$ scattering) 
and by using Eqs.~\eqref{eq:4a} and \eqref{eq:4b} (for $\beta - \alpha\alpha$ scattering). 
The breakup  amplitude is not detailed in the presented  work. However, the dissociation rate is obtained 
through the inelasticity parameter of the elastic amplitude and rearrangement rate, as explained in the following section.

\section{Reaction rates and models}\label{sec3}
The reaction rates are defined in terms of the product of the corresponding cross sections (which are detailed in the
 Appendix~\ref{app:xsection}) and the respective velocity of the colliding particle. By assuming that the atom-dimer 
reaction has initially the particle  $\alpha$ or $\beta$ colliding with the dimer formed by the particles $\alpha\beta$ or $\alpha\alpha$, respectively, 
with energy $E_k\equiv
k^2/(2\mu)$, with $\mu$ being the atom-dimer reduced mass, the total elastic case is such that $\alpha +\alpha\beta\to \alpha +\alpha\beta$
or $\beta +\alpha\alpha\to \beta +\alpha\alpha$. 
Therefore, the elastic reaction rate is given by corresponding product of the elastic cross section with the velocity
$\hbar k/\mu$, such that 
\begin{eqnarray}\label{Kelas}
K_\text{elas}(E_k)=\frac{\hbar k}{\mu}\sigma_\text{el}(E_k).
 \end{eqnarray}
Analogously, the atom-exchange reaction rate can be defined, corresponding to the processes  $\alpha +\alpha\beta\to \beta +\alpha\alpha$ 
(or  $\beta +\alpha\alpha\to \alpha +\alpha\beta$), which is given by 
\begin{eqnarray}\label{Kex}
K_\text{ex}(E_k)=\frac{\hbar k}{\mu}\sigma_\text{ex}(E_k).
\end{eqnarray}
The third possible process refers to dissociation processes
(possible for energies above the three-body continuum),
when for example $\alpha +\alpha\beta\to \alpha +\alpha+\beta $. It is given by
\begin{eqnarray}\label{Kdiss}
K_\text{diss}(E_k)=\frac{\hbar k}{\mu}\sigma_\text{diss}(E_k).
\end{eqnarray}
Given the above, in terms of the corresponding cross-sections, we define the loss-rate coefficient by
 \begin{equation}\label{Kloss}
 K_\text{loss}(E_k)= K_\text{ex}(E_k)+K_\text{diss}(E_k).
 \end{equation}
which refers to the exchange and dissociation, or breakup, cross-sections, respectively.

As in our approach we have two different initial configurations for two-identical particles $\alpha$ with a particle $\beta$, 
the above defined quantities (momentum $k$, energy $E_k$ and reduced mass $\mu$) should be understood such that:
(i) If $\alpha$ is the colliding particle,  $k\equiv k_{\alpha}$, $E_k\equiv E_{k_{\alpha}}$ 
and $\mu\equiv\mu_{\alpha(\alpha\beta)}=m_\alpha(m_\alpha+m_\beta)/(2m_\alpha+m_\beta)$;
(ii) if $\beta$ is the colliding particle,  $k\equiv k_{\beta}$, $E_k\equiv E_{k_{\beta}}$ 
and $\mu\equiv\mu_{\beta(\alpha\alpha)}=(2m_\beta m_\alpha)/(2m_\alpha+m_\beta)$.

The corresponding partial wave $\ell$ decomposition of the above reaction rates [Eqs.~\eqref{Kelas} - \eqref{Kloss}] 
can be expressed in terms of the non-diagonal S-matrix, for the elastic and for the atom-exchange channel.
Given that $K_\text{x}(E_k)\equiv K_\text{x}$, for $\text{x}=$ (elas, ex, diss, loss), we have 
the following:
{\small
\begin{eqnarray}
&& K_\text{elas}=\frac{\pi\hbar}{\mu k}\sum_{\ell=0}^\infty (2\ell+1)\mid 1-S^\ell_\text{el}(E_k)\mid^ 2, \label{Kelasl}\\
&& K_\text{ex}=\frac{\pi\hbar}{\mu k}\sum_{\ell=0}^\infty (2\ell+1) \mid S^\ell_\text{ex}(E_k)\mid^ 2, \label{Kexl}\\
&& K_\text{diss}=\frac{\pi\hbar}{\mu k}\sum_{\ell=0}^\infty (2\ell+1) (1-\mid S^\ell_\text{el}(E_k)\mid^ 2-
 \mid S^\ell_\text{ex}\mid^ 2), \label{Kdiss2}\\
&&K_\text{loss}= \frac{\pi\hbar}{\mu k}\sum_{\ell=0}^\infty (2\ell+1) (1-\mid S^\ell_\text{el}(E_k)\mid^ 2).
\label{Klossl} 
\end{eqnarray}
Our aim is to study  the reaction rates for the atom-dimer collisions, in which 
$\alpha\equiv ^4$He, with $\beta\equiv ^6$Li, $^7$Li, and $^{23}$Na, by considering all possible combinations
for the three $\alpha\alpha\beta$ systems. 
The relevant information about these systems are provided in Tables~\ref{tab1} and \ref{tab2}, with the 
dimer and trimer binding energies being obtained from Refs.~\cite{2017Suno} (a1)  and \cite{1998Yuan} (a2).
The respective configuration-space behaviors of the two-body potentials $V(r)$, which were used to obtain the dimer  
energies provided in Refs.~\cite{1998Yuan} and \cite{2017Suno},  are represented in Fig.~\ref{fig01}.  

The corresponding two- and three-body ground-state binding energies (absolute 
values, given in mK) are presented in Table~\ref{tab1}, in which we also include the given excited three-body binding 
energies, whenever known.  More specifically, the results for (a1)  and (a2) were obtained by considering interactions 
 from 
 Refs.~\cite{1999Kleinekathofer} and \cite{1996KTTY}, respectively, as shown in Fig.~\ref{fig01}. 
Other model calculations exist for the systems we are studying, beyond the selected (a1) and (a2) models, as described 
in Ref.~\cite{2017Suno}. They are from Ref.~\cite{2002Gianturco}, with $\alpha\alpha$ and $\alpha\beta$ potentials given
by Refs.~\cite{1997Kleinekathofer,1999Kleinekathofer,1995Tang}; from Ref. \cite{2002Gianturco,2000Gianturco}, with potentials 
given by Ref.~\cite{1991Aziz,1994Cveko}; from Ref. \cite{2014Wu}, with $\alpha\alpha$ potential given by 
Ref.~\cite{1991Aziz}, and $\alpha\beta$ by Ref.~\cite{1999Kleinekathofer}. However, we select only (a1) and (a2), 
among the ones which are discussed in Refs.~\cite{2017Suno,2018ShalchiPRA98}, as they are providing values for all 
the cases we are considering, having binding energies enough distinct for a significant comparative study.

One should
notice that the binding-energy difference between the dimers $^4$He$-^6$Li and $^4$He$-^7$Li, shown in Table~\ref{tab1},
refers to the corresponding isotopic mass difference. More relevant to be noticed in the table, for such weakly-bound 
systems, is the binding energy sensibility on the potential depth differences, as verified between the (a1) and (a2) models
for the He-Li system (see the inset of Fig.~\ref{fig01}), which changes the kind of the 
atom-exchange reaction (endothermic or exothermic).

The Table~\ref{tab2} refers to the kind of each reaction channel considered in the two potential models we have used (ZR and FR) 
with both different inputs.
In all the cases under our analysis, in which we have two $^4$He atoms, identified by $\alpha$ and a third atom 
identified by $\beta$=($^6$Li, $^7$Li, $^{23}$Na), a weakly bound molecule exists, such that 
six entrance channels are possible for the atom-dimer reactions, which are the following:\\
 \begin{equation}\left.\begin{array}{lcll}
(1) & ^{4}{\rm He}  &+& ^4{\rm He}^6{\rm Li},\\
(2) & ^{6}{\rm Li}    &+& ^4{\rm He}_2,\\
(3) & ^{4}{\rm He}  &+& ^4{\rm He}^7{\rm Li},\\
(4) & ^{7}{\rm Li}    &+& ^4{\rm He}_ 2,\\
(5) & ^{4}{\rm He}   &+& ^4{\rm He}^{23}{\rm Na},\\
(6) & ^{23}{\rm Na} &+& ^4{\rm He}_2 .\end{array}
\right\}\\ \label{channels} \end{equation}
For each one of these atom-dimer initial reaction channels, we can obtain the elastic, atom exchange and dissociation reaction rates. 
In order to compute these quantities, among the available potential models which were investigated in 
Refs.~\cite{2017Suno,1998Yuan}, we select two of them, as we already mentioned, which provide less-similar values for the dimer and
trimer binding energies that we are considering.
 These potential models will be used to provide the 
necessary binding energies for the inputs to adjust the parameters of our zero-range  and finite-range $s-$wave 
separable interactions, within a Faddeev three-body formalism for the atom-dimer collision (for details, see sect. \ref{subsec:atomdimercollision}). 

\begin{figure}[thb]
\includegraphics[width=8cm]{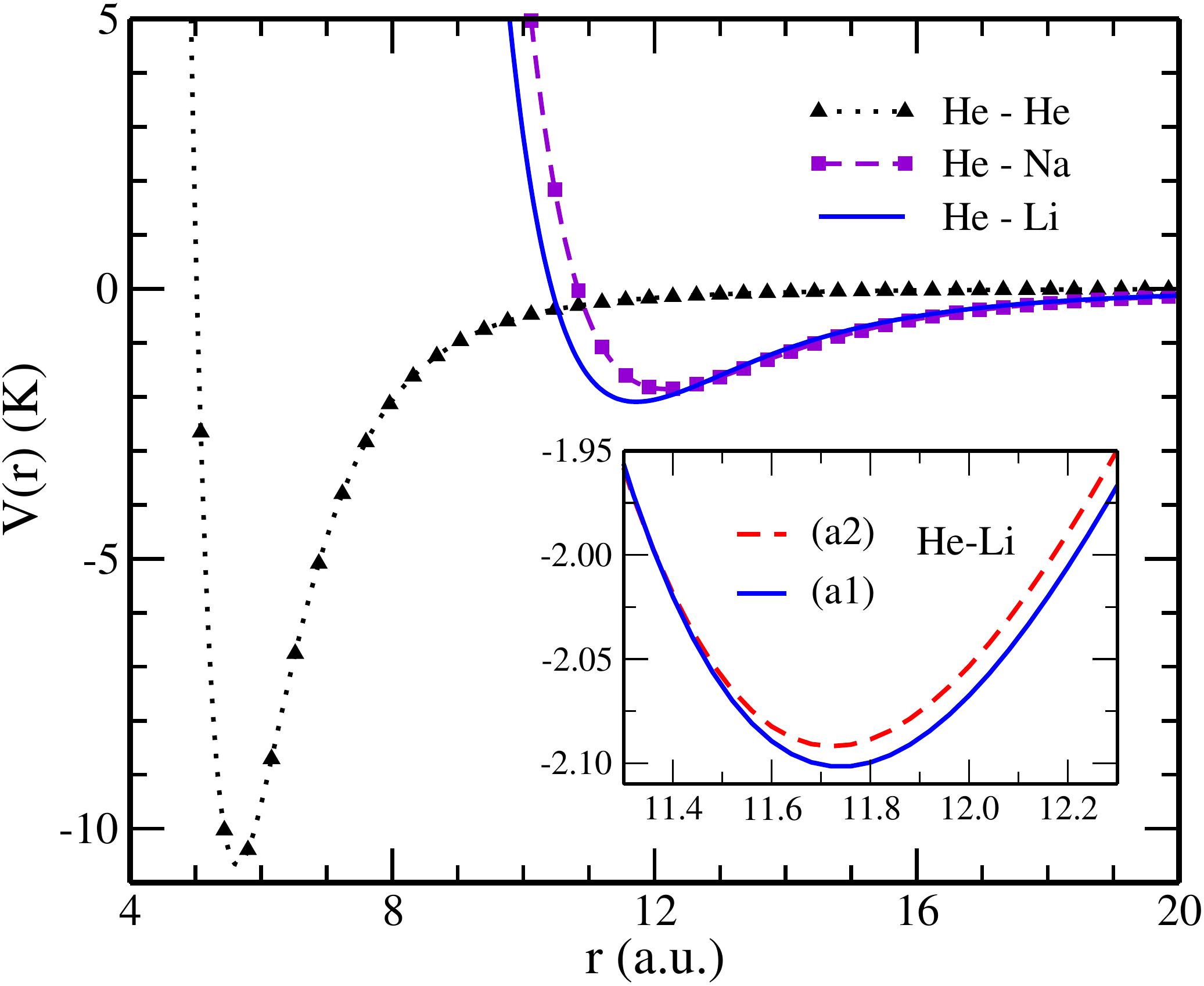}
\vspace{-.3cm}
\caption{
Dimer potentials for the He-He (dotted-line with triangles), He-Na (dashed-line with squares)  and He-Li (solid line)
systems, used in the three-body calculations of~\cite{2017Suno} (a1) and ~\cite{1998Yuan} (a2), are shown in the main frame.
The He-Li (a1) and (a2) potentials,  
respectively derived in Refs.~\cite{1999Kleinekathofer} and \cite{1996KTTY},  
differ by tenth's of  mK, being indistinguishable and represented by a single line 
in the main frame. They are shown in the inset by the solid  (a1) and dashed (a2) lines. 
}
\vspace{-0.5cm}
\label{fig01}
\end{figure}

\begin{table}[tbh!]
  \caption{This table provides the available dimer and trimer binding energies (in mK),  combining two  
  $^4$He with $^6$Li, $^7$Li and $^{23}$Na atoms. The given results for (a1) are from 
  Ref.~\cite{2017Suno} and for (a2) from \cite{1998Yuan}.  
With (*) we have the available excited binding energies, from Ref.~\cite{2017Suno}.}
\begin{tabular}{lcccc}
\hline\hline
Molecule      &&(a1) 	&&(a2) \\
\hline\hline
$^4$He$_2$ 	&\;\;\;&1.31	&\;\;\;&1.31	\\
$^4$He$^6$Li	&&1.515	&&0.12	\\
$^4$He$^7$Li &&5.622	&&2.16	\\
$^4$He$^{23}$Na	&&28.98	&&28.98	\\
\hline \hline
$^4$He$_2-^{6}$Li	&&57.23	&&31.4	\\
($^4$He$_2-^{6}$Li)$^*$	&&1.937	&& -	\\
$^4$He$_2-^{7}$Li	&&79.36	&&45.7	\\
($^4$He$_2-^{7}$Li)$^*$ &&5.642&& -	\\
$^4$He$_2-^{23}$Na&&150.9	&&103.1	\\
 \hline\hline
  \end{tabular}
  \label{tab1}
 \end{table}

\begin{table}[tbh!]
\caption{Classification of reaction channels, identified in \eqref{channels}, 
as endothermic or exothermic according to dimer binding energies of models
(a1) and (a2), given in Table~\ref{tab1}, for elastic and exchange processes.
For dissociation ($\to ^{4}$He+$^{4}$He+$^{6}$Li), all channels are endothermic.}
\begin{tabular}{ccccc} 
\hline\hline
Channel&Model& Elastic & & Exchange \\
\hline\hline
(1)&&$^{4}$He + $^4$He$^6$Li & & $^{4}$He$_2$ +$^{6}$Li   \\
&(a1)&                                   & & endothermic  \\
&(a2)&                                   & & exothermic   \\
\hline
(2) && $^{6}$Li + $^{4}$He$_{2}$ & & $^{4}$He + $^4$He$^6$Li  \\ 
&(a1) &                                   & & exothermic  \\
&(a2) &                                  & & endothermic  \\
\hline
(3) && $^{4}$He + $^4$He$^7$Li & & $^{4}$He$_2$ +$^{7}$Li    \\
&(a1) &                                   & & endothermic   \\
&(a2) &                                   & & endothermic   \\
\hline
(4) && $^{7}$Li + $^{4}$He$_2$ & & $^{4}$He + $^4$He$^7$Li \\
&(a1)&                                   & & exothermic   \\
&(a2)&                                   & & exothermic    \\
\hline 
(5) && $^{4}$He + $^4$He$^{23}$Na & & $^{4}$He$_2$ + $^{23}$Na \\ 
&(a1)&                                   & & endothermic   \\
&(a2)&                                   & & endothermic   \\
\hline
(6)  && $^{23}$Na + $^{4}$He$_2$  & & $^{4}$He + $^4$He$^{23}$Na \\
&(a1)&                                   & & exothermic  \\
&(a2)&                                   & &exothermic  \\
\hline
\hline
  \end{tabular}
  \label{tab2}
 \end{table}

In order to compute the reaction rates, as mentioned before,
for each set of binding energies (a1) and (a2) provided in Table~\ref{tab1},  
we have adjusted our zero-range and finite-range two-body interactions, for 
which the corresponding kernels of the scattering equations are detailed in Appendix~\ref{app:kernels}.
In the zero range model, we have the two-body amplitude parametrized by the diatomic binding energies, 
together with a regularizing momentum parameter fixed by the triatomic molecule.
For the finite-range interaction, we  assume a rank-one separable Yamaguchi potential, given by
\begin{eqnarray}\label{YamguchiV}
 V_{ij}(p,p')=\lambda_{ij}\, \frac{1}{p^2+\gamma_{ij}^2}\, \frac{1}{p'^2+\gamma_{ij}^2},
\end{eqnarray}
where $ij=\alpha\alpha$ or $\alpha\beta$, respectively, for the $\alpha\alpha$ or $\alpha\beta$ two-body subsystems.
$\lambda_{ij}$ and $\gamma_{ij}$ refer to the strengths and ranges of the respective two-body interactions.  
As in the present approach we consider only bound (negative) two-body subsystems, $E_{ij}=-B_{ij}$, the corresponding 
relations for the strengths and ranges are given by
\begin{eqnarray}
\lambda^{-1}_{ij}=\frac{-2\pi\mu_{ij}}{\gamma_{ij}(\gamma_{ij}+\kappa_{ij})^2},\;\;\;
 r_{ij}=\frac{1}{\gamma_{ij}}+\frac{2\gamma_{ij}}{(\gamma_{ij}+\kappa_{ij})^2},
\end{eqnarray}
where $r_{ij}$ are the effective ranges, with 
\begin{eqnarray}
   \kappa_{\alpha\alpha}&\equiv&\sqrt{-2\mu_{\alpha\alpha} E_{\alpha\alpha}},
   \;\;\; \kappa_{\alpha\beta}\equiv\sqrt{-2\mu_{\alpha\beta}
   E_{\alpha\beta}}
\, .\end{eqnarray}

For the case of FR potential given in Eq.~\eqref{YamguchiV}, the parameters with corresponding ranges and scattering 
lengths, 
are shown in Table~\ref{tab3}, given in three blocks 
for the cases with ($^4$He$^7$Li), ($^4$He$^6$Li) and ($^4$He$^{23}$Na). We observe that, in all the cases, for the
dimer $^4$He$_2$ binding energy, the accepted value $B_{\alpha\alpha}=$1.31 mK is being considered, with the
corresponding parameters given in this table.

 \begin{table}[tbh!]
\caption{Parameters used in the FR $s-$wave separable interaction, with the corresponding ranges and scattering lengths, 
in order to reproduce the respective binding energies given in Table~\ref{tab1} for the 
model potentials (a1) and (a2).}
\begin{tabular}{ccccc} 
\hline\hline
Dimer&Model&$\gamma_{\alpha\beta}$(\AA$^{-1}$)&$r_{\alpha\beta}$(\AA)&$a_{\alpha\beta}$(\AA)\\
\hline\hline
{\bf$^4$He$^6$Li}&(a1) & 0.17&   15.85&    90.38\\
&(a2) & 0.14&    20.04&       300.37\\
 \hline
 {\bf$^4$He$^7$Li}& (a1)& 0.17& 14.77& 50.08\\
&(a2)& 0.14 &       19.02 &    77.43\\
 \hline
{\bf$^4$He$^{23}$Na}&(a1)& 0.16  &      12.44   &     25.34\\   
&(a2)& 0.09 &  19.0  &  34.24\\
 \hline
{\bf$^4$He$_2$}&(a1)\&(a2)&  0.39  &    7.34     & 100   \\   
 \hline\hline
  \end{tabular}
  \label{tab3}
 \end{table}

The results for the reaction rates are organized according to the Table~\ref{tab2} and  are calculated with the
ZR and FR potential models,  in order to exhibit the model independent
features of the present results, for up to the $d-$wave contribution. The choice of the potential, ZR or FR, affects mainly the
$s-$wave contribution, which is more sensitive to the effective range. 
Reminding that the models fit the same diatomic and 
triatomic binding energies, we found that the bulk results are almost model independent; particularly, 
for the higher partial waves the results are 
to a large extend universal. Such features gives 
robust outcomes of our model calculations, as we will present in the following. However, the reaction rates are
bounded by the binding energies provided  in Refs.~\cite{2017Suno} and
 \cite{1998Yuan}, while our predictions allow to discriminate between
the very different values given for the model (a1) and  (a2). We perform calculations up to kinetic energies of 0.1 K.
Note also that, we have carefully checked that almost no deviation from the unitarity appears in our numerical  solutions
of the scattering equations, for all the colliding energies being considered.  In particular,  we verified that the coupled 
elastic and atom-exchange channels  S-matrix  is unitary below the dissociation threshold,   as obtained numerically for
both zero- and finite-range interaction models.

\begin{table}[thb!]
  \caption{
For the molecules identified in the first column, in the 2nd, 3rd and 4th columns 
we have the corresponding two-body energies (mK), scattering lengths (\AA), and trimer energies (mK) 
(obtained from Ref.~\cite{2017Suno}), respectively. In the 5th to 7th columns we have the first excited 
bound-state energies (in mK).
Results obtained from Refs.~\cite{2017Suno} and \cite{2018ShalchiPRA98}, as indicated in the last row.
In addition, for the excited states, we show results obtained by using zero-range (ZR) and finite-range (FR) 
one-term separable interactions.  The parameters for the $^4$He-$^4$He subsystem are given in 
Tables~\ref{tab1} and \ref{tab3}. 
}
{\footnotesize \begin{tabular}{c|ccc|ccc}
 \hline\hline
molecule&
He-Li&
$a_{\text{He}\text{Li}}$&
${\text{He}_2\text{Li}}$&
$({\text{He}_2\text{Li}})^*$&
$(\text{He}_2\text{Li})^*$ &
$(\text{He}_2\text{Li})^*$
\\
\hline\hline
 $^4$He$_2$-$^6$Li &1.515&100&57.23        &1.937&1.901&1.977\\
$^4$He$_2$-$^7$Li &5.622&48.84&79.36     &5.642&-&5.672\\
Ref.     &\cite{2017Suno} & \cite{2018ShalchiPRA98} &  \cite{2017Suno}&  \cite{2017Suno}   
&(ZR)~\cite{2018ShalchiPRA98}  &(FR)~\cite{2018ShalchiPRA98}\\
 \hline\hline
  \end{tabular}
  }
  \label{tab4}
 \end{table}

Before closing this section, it is necessary to point out that 
the molecules  $^4$He$_2\,^6$Li and  $^4$He$_2\,^7$Li present excited Efimov states close to the lowest 
scattering thresholds as verified in Ref.~\cite{2017Suno} and for the ZR and FR 
potentials in \cite{2018ShalchiPRA98}. These results corresponding to model (a1) are provided in Table~\ref{tab4}, 
where in \cite{2018ShalchiPRA98} slightly different scattering lengths were used with respect to Table~\ref{tab3}, 
which is not relevant. It is important to note that weakly bound triatomic states close to the threshold affect the
reaction rates at low energies.

In what follows we will present our results for the atom-dimer reaction rates by considering the 
possible three-atom systems $\alpha\alpha\beta$ with the particle $\alpha$ being $^4$He and 
$\beta$ being one of the species $^{6}$Li, $^{7}$Li or $^{23}$Na. We split the presentation in the next 
three sections: In Sect.~\ref{sec4} we consider $\beta\equiv ^{6}$Li;
in Sect.~\ref{sec5}, $\beta\equiv ^{7}$Li; and in Sect.~\ref{sec6}, $\beta\equiv ^{23}$Na. In each of these three
sections, we have two subsections for the reaction-rate results, such that  
$\alpha + \alpha\beta$ reactions are presented in (A), with the 
$\beta + \alpha\alpha$ reactions presented in  (B), with all the possibilities being presented in Table~\ref{tab2}.

\section{Three body reactions with Helium-4 and Lithium-6}\label{sec4}
\subsection{Reaction rates for $^{4}$He + $^4$He$^6$Li}
The calculations  for the reaction rates, i.e., elastic, exchange and loss, for the $^{4}$He + $^4$He$^6$Li collision 
 are  shown in the following. The parameters of the zero-range and 
rank-1 separable $s-$wave potential (see Table \ref{tab2}) models are fitted to reproduce the binding energies  
obtained from the choices of potentials models given in Table \ref{tab1}.  The results are presented
in Fig.~\ref{fig02} for model (a1) and in \ref{fig03} for model (a2), and in each figure the outcome of the
zero-range and separable finite range potential are shown. 

\begin{figure}[thb]
\includegraphics[width=8cm]{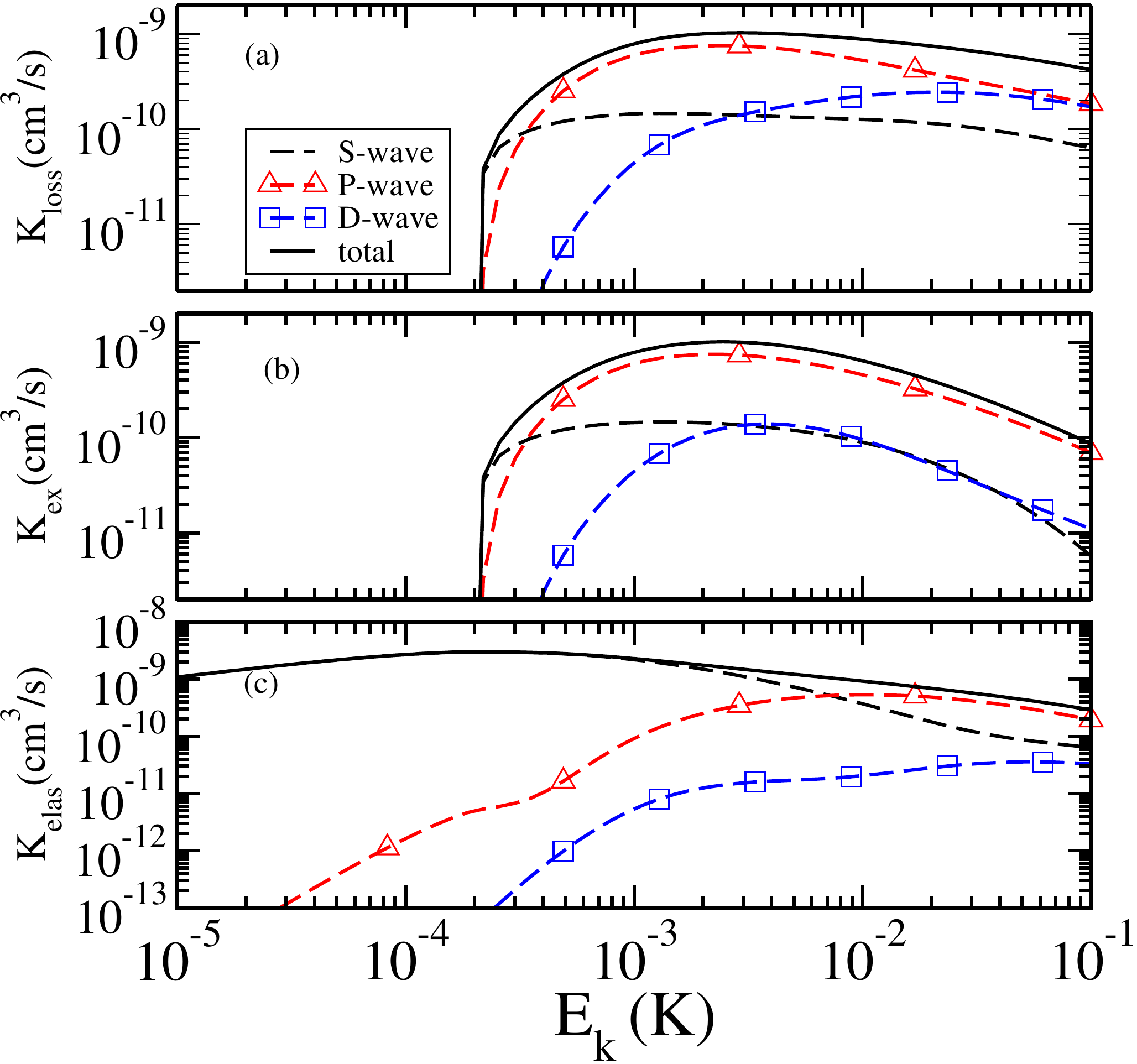}
\includegraphics[width=8cm]{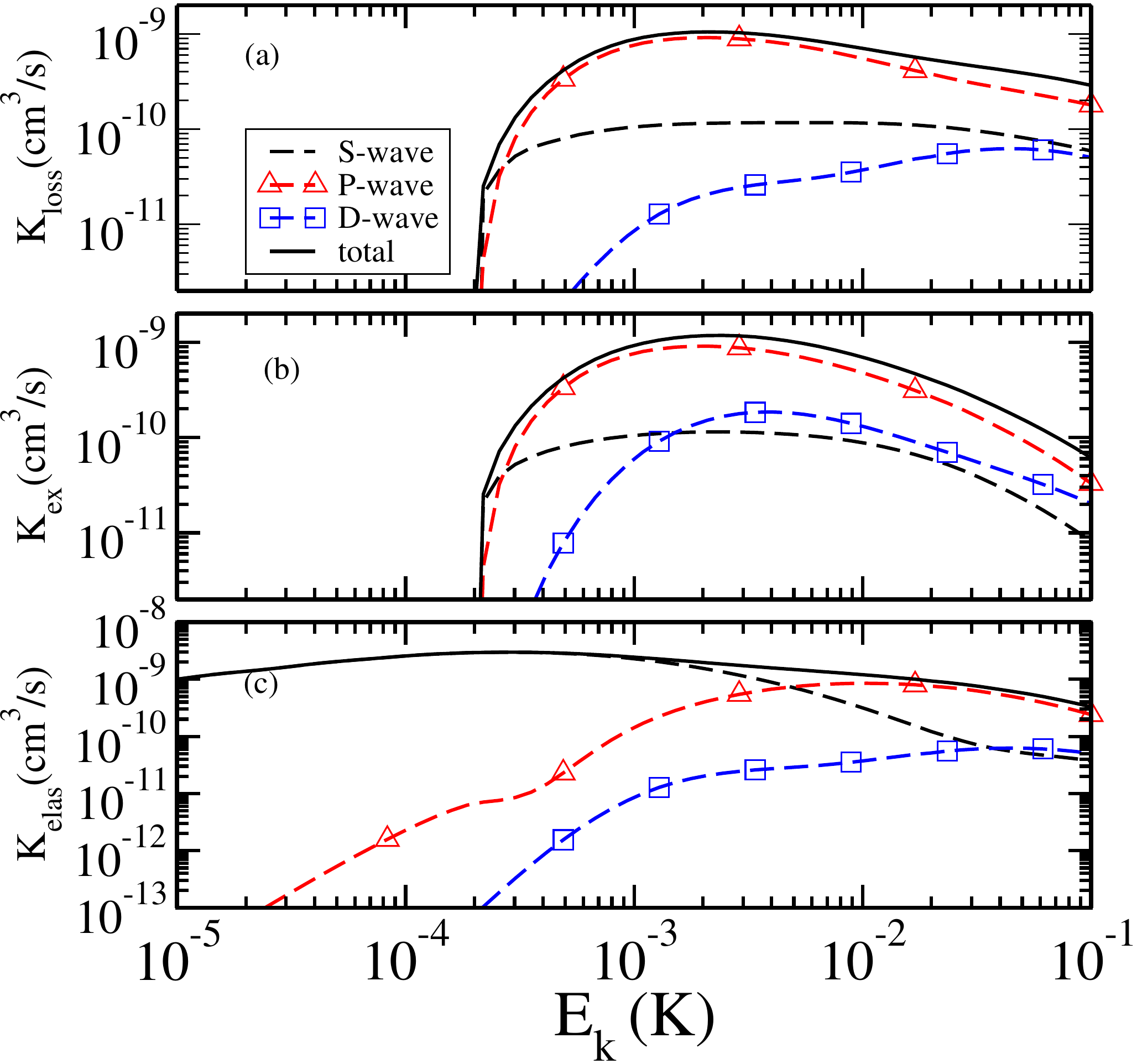}
\vspace{-.3cm}
\caption{
Reaction rates for $^{4}$He + $^4$He$^6$Li, using zero-range (upper-set of panels) and 
finite-range (lower-set of panels) interactions, considering the (a1) model.
}
\vspace{-0.5cm}
\label{fig02}
\end{figure}

The exchange and loss rates for the $^{4}$He + $^4$He$^6$Li with model (a1), shown in Fig.~\ref{fig02}, correspond to  
endothermic reactions:  the threshold energy in order to open the exchange channel is 0.2 mK, with the three-body dissociation 
 at 1.5 mK. The $s-$wave elastic rate is dominant below opening the threshold attaining values as larger as
10$ ^{-9}$cm$^3$/s, for both the zero-range and finite-range potentials. Once the  atom exchange  channel opens,
i.e. $^{4}$He + $^4$He$^6$Li$\to$$^{4}$He$_2$ + $^{6}$Li, the $p-$wave becomes noticeable and right away 
gives the major contribution to both the atom exchange and loss rate, even above the dissociation channel. 

\begin{figure}[thb]
\includegraphics[width=8cm]{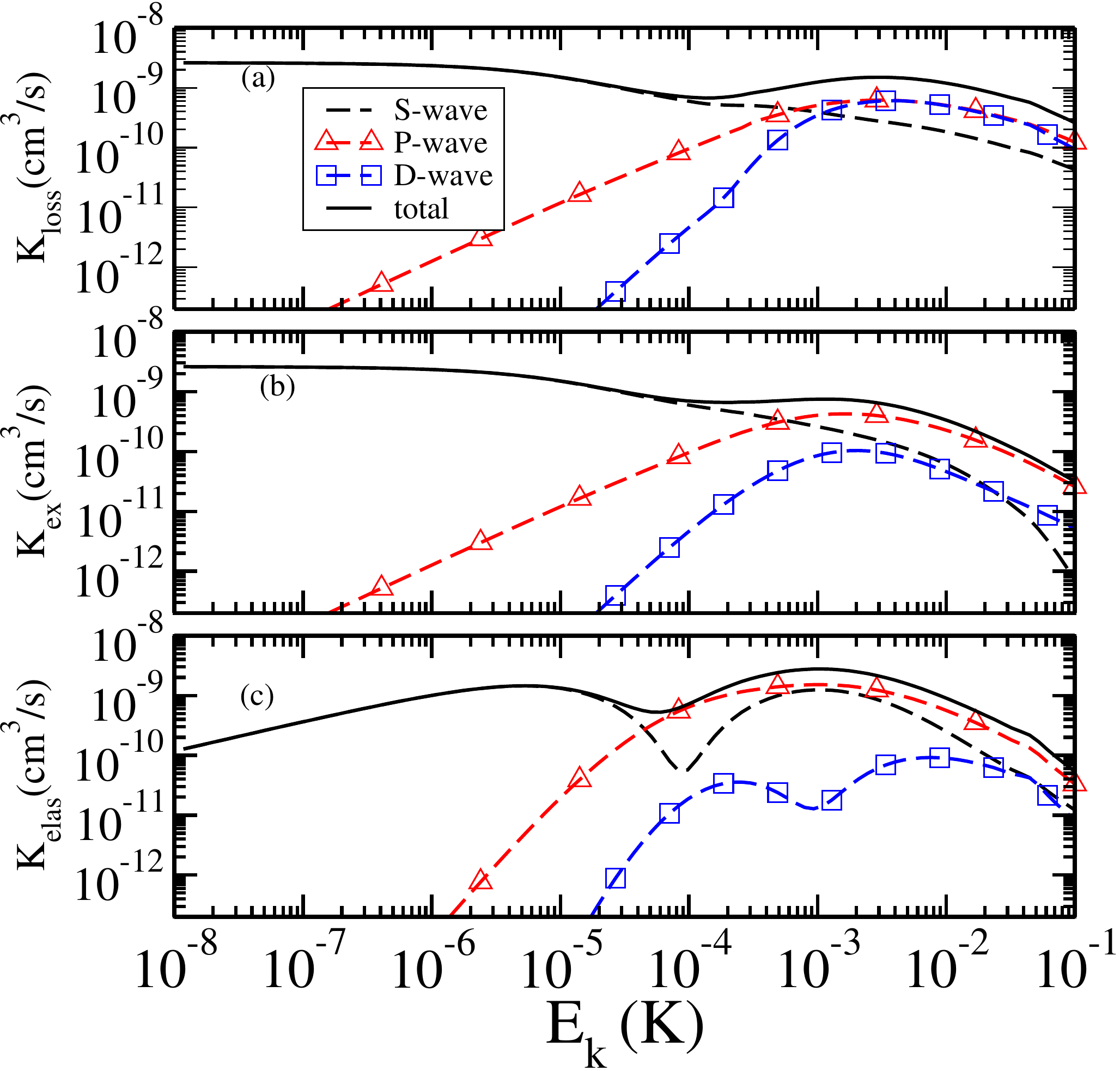}
\includegraphics[width=8cm]{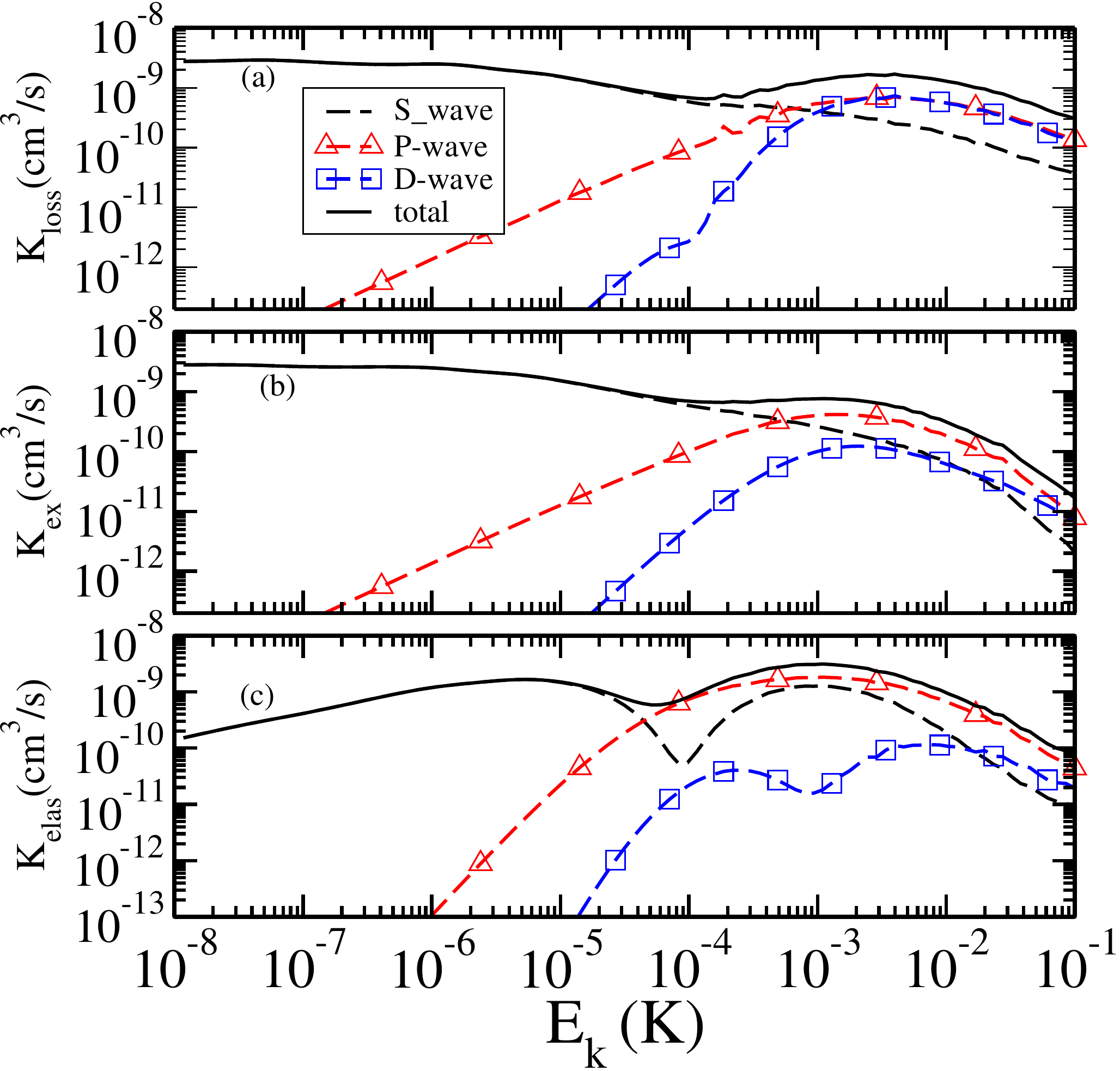}
\vspace{-.3cm}
\caption{
Reaction rates for $^{4}$He + $^4$He$^6$Li, using zero-range (upper-set of panels) and 
finite-range (lower-set of panels) interactions, considering the (a2) model.
}
\vspace{-0.5cm}
\label{fig03}
\end{figure}

The $p-$wave dominance comes from the well-known one-atom exchange diagram, which due to the involved small binding 
energy has a pole close to the scattering threshold. Physically in the exchange process,  the  incoming $^4$He  picks 
the other one from  the $^4$He$^6$Li molecule, and the remaining $^{6}$Li is moving backwards with respect to the 
incoming  $^4$He, enhancing the $p-$wave  contribution to the reaction process.

The $s-$wave projection of the one-atom exchange diagram is also associated with the appearance of the Efimov effect, as 
such  diagram comes also as the kernel of the integral equations for the scattering and bound state, both for the 
zero-range and separable potential models.  One can observe that both models predict very similar reaction rates, as these 
low-energy process are quite universal, and determined by the two-atoms low energy observables (in this case the binding 
energies) and one three-body input, as the binding energies of the triatomic molecules. The separable interaction model 
has indeed scattering lengths almost three times the effective ranges (see Table~\ref{tab2}), characterizing well a short range 
potential with three-body low energy observables weakly dependent on the range.
It is important to emphasize that the $s-$wave observables are  more sensitive to the potential, as they require for the limit 
of zero-range interaction the information of the triatomic binding energy, while the $p$, $d$, .... waves are essentially sensitive to 
the on-shell low energy two-atom amplitude. 

The reaction rates for  $^{4}$He + ($^4$He$^6$Li) obtained with model (a2) are shown in  Fig.~\ref{fig03}. Now, 
the atom exchange reaction is exothermic (see Table~\ref{tab3}), with the three-atom continuum opening at 0.12mK. The 
$s-$wave is dominant up to this threshold with rates above 10$ ^{-9}$cm$^3$/s. The  $s-$wave elastic rate has a minimum 
at the inelastic threshold, with the $p-$wave emerging as the dominant one above it. The $d-$wave is in general less important,
but can be comparable with the $s-$wave for higher energies.
The manifestation of the minimum in the $s-$wave can be clearly seen in the elastic rate.
The $d-$wave phase shift has also a zero around 1mK, reflected as a minimum of the elastic amplitude, due to the absorption.
This feature is model independent, as one can observe by comparing the two potential models. 
The atom exchange rate is dominated by the $p-$wave above the dissociation threshold, with the $s-$ and $d-$waves  
representing less than 20\% of the rate.
The calculated loss rate for the reaction $^{4}$He + $^4$He$^6$Li  is also shown in  Fig.~\ref{fig03} for the model (a2).  
A noticeable dip appears at 0.1 mK, which comes from the fast increase of the $p-$ and $d-$waves contribution to the 
dissociation  channel, while the $s-$wave becomes less relevant.  The dissociation  channel is more efficiently populated 
by the higher partial waves, as the relevant inputs are only the binding energies of the $^4$He dimer and the $^4$He$^6$Li, 
 which in this case are pretty small.

 As already mentioned, beyond $s-$wave the three-body scattering amplitudes are dominated by the on-shell two-atom  low-energy 
 T-matrix, which sizes the kernel of the three-body scattering equation as can be clearly verified, for example, in the zero-range model. 
In this way,  the bound state pole of the very weakly bound $^4$He$^6$Li  molecule  enhances even more the kernel and together
the contribution of the $p-$ and $d-$waves to the loss rate. 
 The elastic, 
 atom exchange and loss rates clearly distinguish the models (a1) and (a2), providing a mean to indirectly access the on-shell 
 quantities. The difference in the binding energy of the $^{4}$He$_2$$^{6}$Li molecule  in models (a1) and (a2) is relevant for the $s-$wave, while it is barely 
 perceived by the $p-$ and  $d-$waves, that show  sensitivity to the different values of the $^{4}$He$^{6}$Li energy even above the dissociation threshold.

\subsection{Reaction rates for  $^{6}$Li + $^{4}$He$_2$}\label{subsec6LiHe}
The  results for the reaction rates in the $^{6}$Li + $^{4}$He$_2$ collision are shown in
Figs. \ref{fig04} and \ref{fig05}, with the $^{6}$Li + $^{4}$He$_2$ $\to\, ^4$He + $^{4}$He$^{6}$Li reaction
being exothermic for model (a1) and endothermic for model (a2) (see Table \ref{tab3}). 
For the potential model (a1),  the dissociation channel opens at 1.3 mK, dominating the losses only above 10 mK,  
as one can observe from the corresponding panels of Fig.~\ref{fig04}.
The elastic rate is dominated by the $s-$wave up to 10 mK, when the $p-$wave takes over. The atom 
exchange rate, already discussed, has a major contribution from the $p-$wave, which raises above 
the $s-$wave around 0.1 mK, while the $d-$wave up to 0.1 K is not relevant. 

\begin{figure}[thb]
\includegraphics[width=8cm]{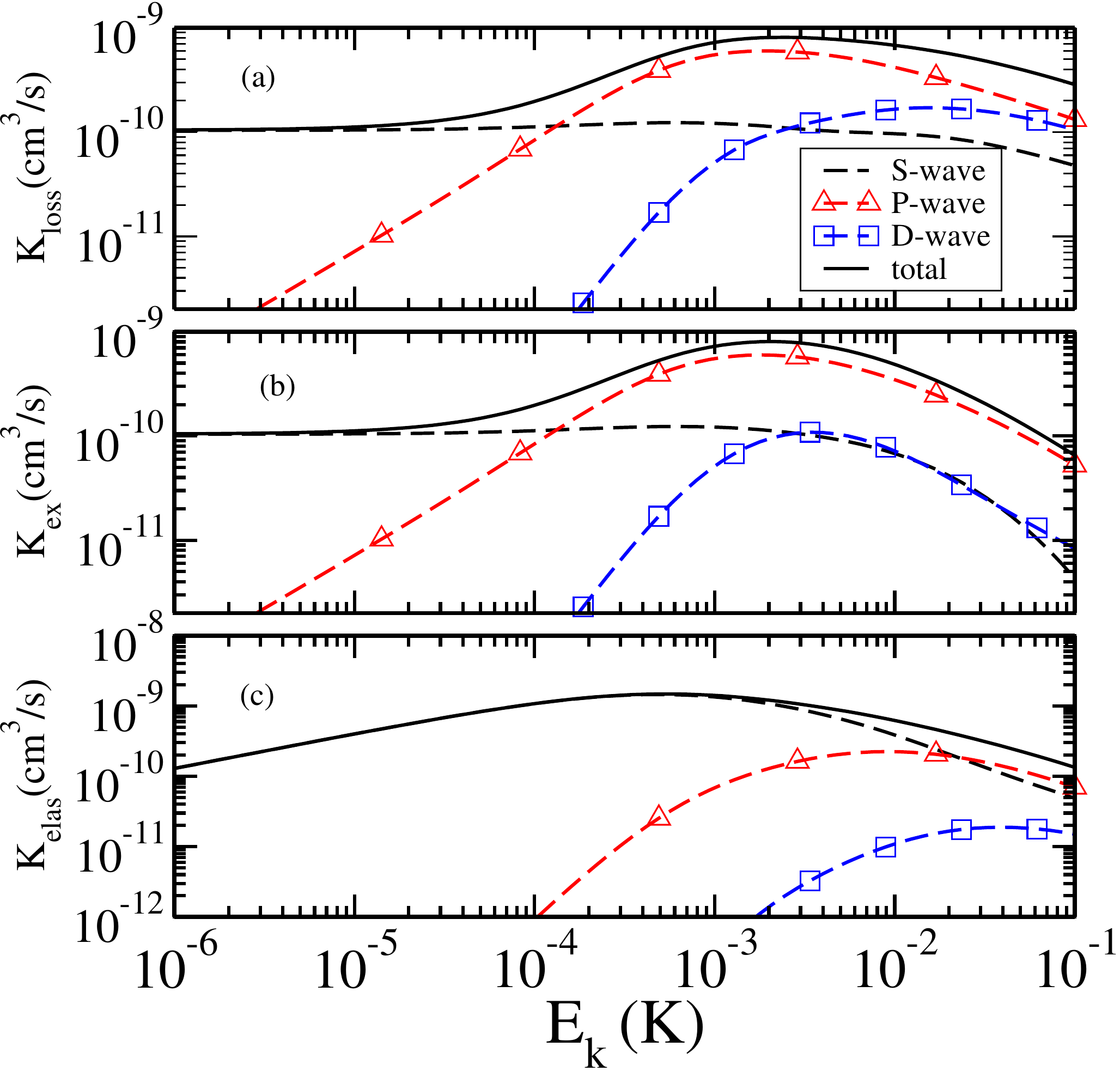}
\includegraphics[width=8cm]{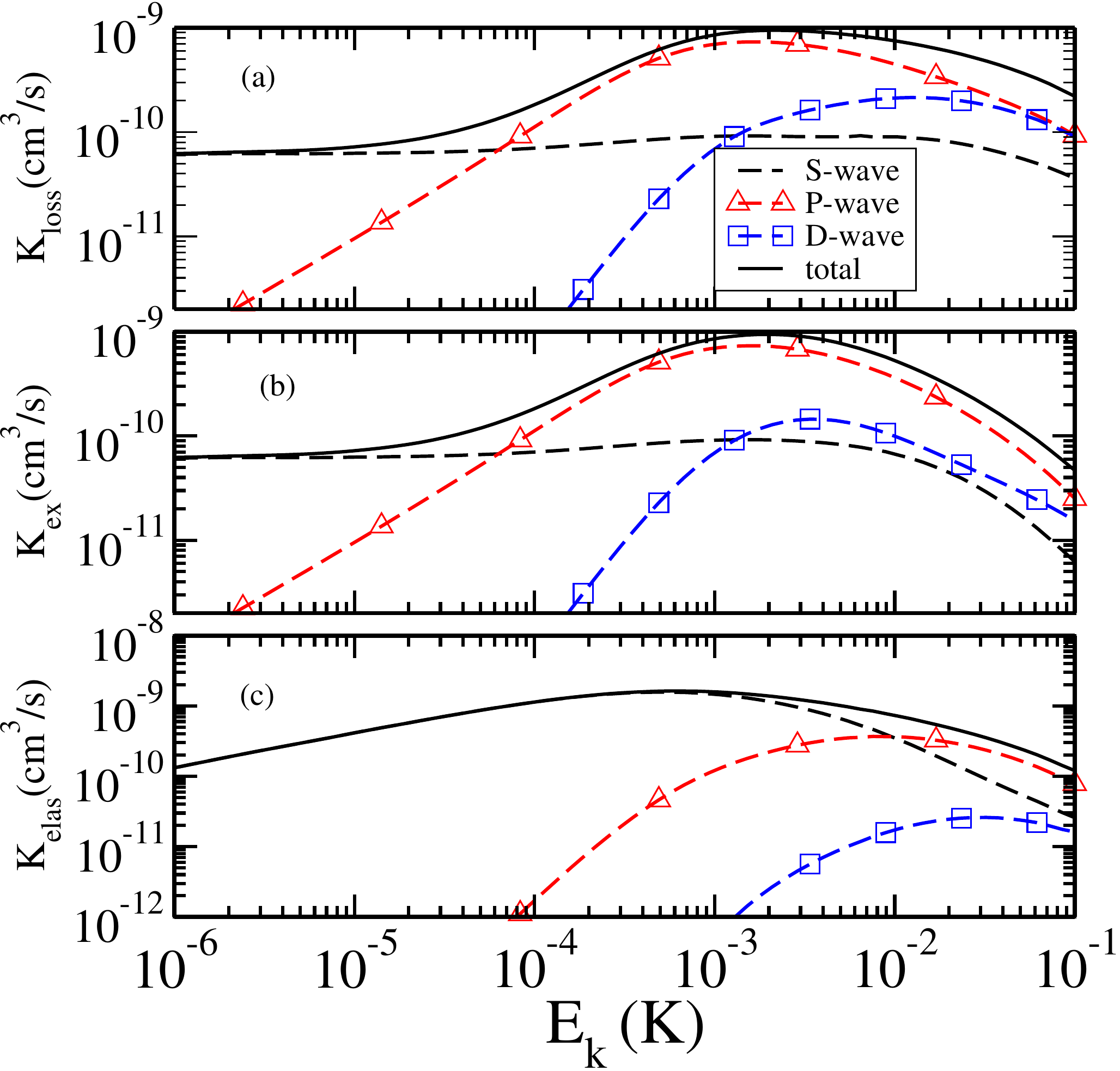}
\vspace{-.3cm}
\caption{
Reaction rates for $^{6}$Li + $^{4}$He$_2$, using zero-range (upper-set of panels) and 
finite-range (lower-set of panels) interactions, considering the (a1) model.
}
\vspace{-0.2cm}
\label{fig04}
\end{figure}

The dissociation component of the loss rate has an overwhelming contribution from the $p-$wave up to about 10 mK, when 
the $d-$wave becomes competitive due to its contribution, essentially, to the dissociation process. We have to remind that 
the $p-$ and $d-$waves are determined by the bindings energies of the diatomic molecules  $^4$He$_2$ and  the very 
weakly bound   $^4$He$ ^6$Li, which gives a long range tail for the attractive Efimov like potential~\cite{Efimov},  that of 
course damps the centrifugal  barrier enhancing the importance of the higher partial waves, even  at low energies. The relative relevance of the 
$d-$wave contribution can be clearly seen in the loss 
rate.

Comparing the calculations of the zero- and finite-range potential models, for the same inputs from the set (a1), it is clear the  
independence of the bulk results on the detail of the potential beyond the dimer and trimer binding energies. Both 
models exhibit a strong enhancement of the atom exchange and loss rates between 2-3 mK, increasing up to 
10$^{-9}$cm$^3$/s,  with the he $p-$wave playing a major role.

\begin{figure}[thb]
\includegraphics[width=8cm]{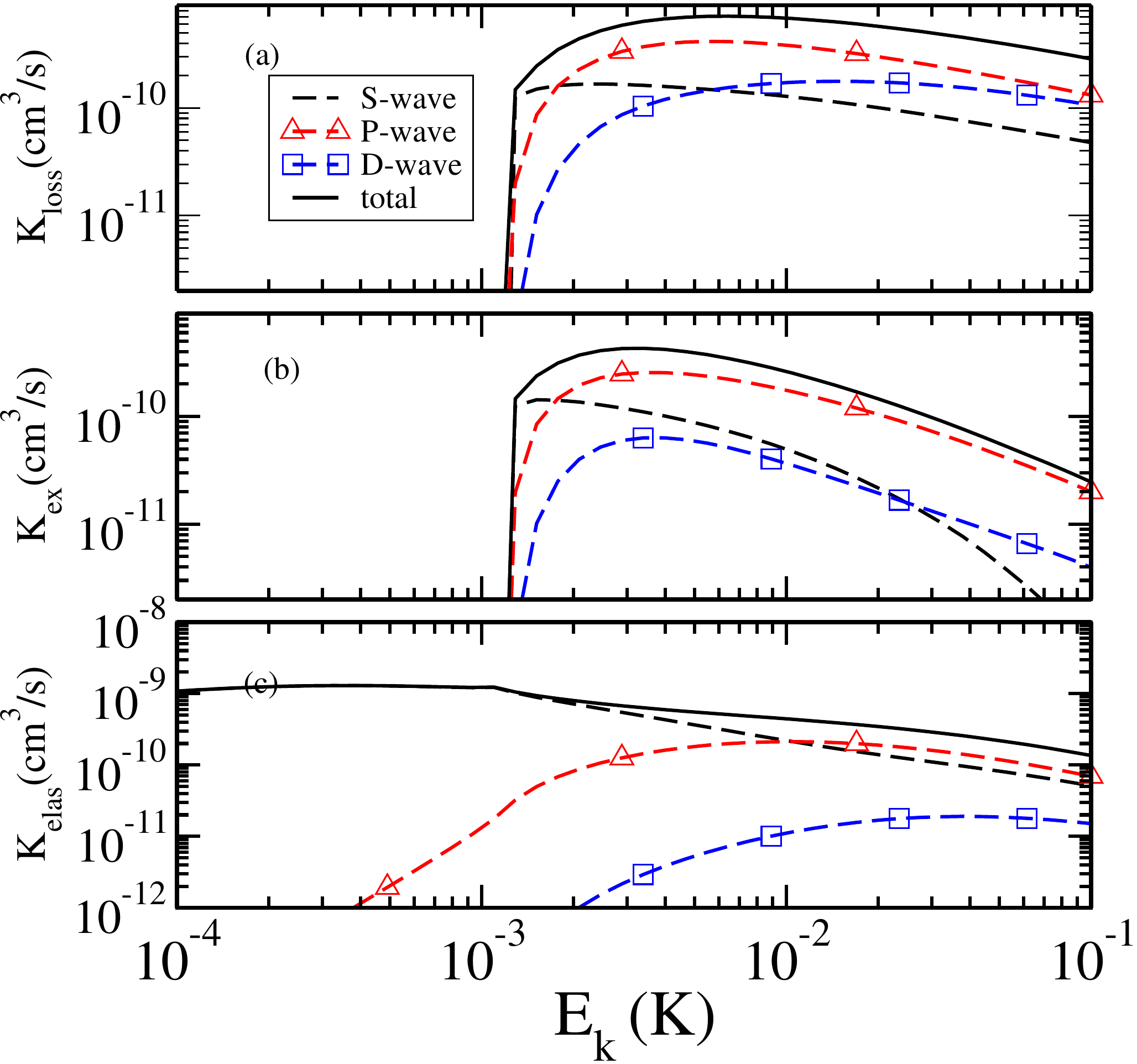}
\includegraphics[width=8cm]{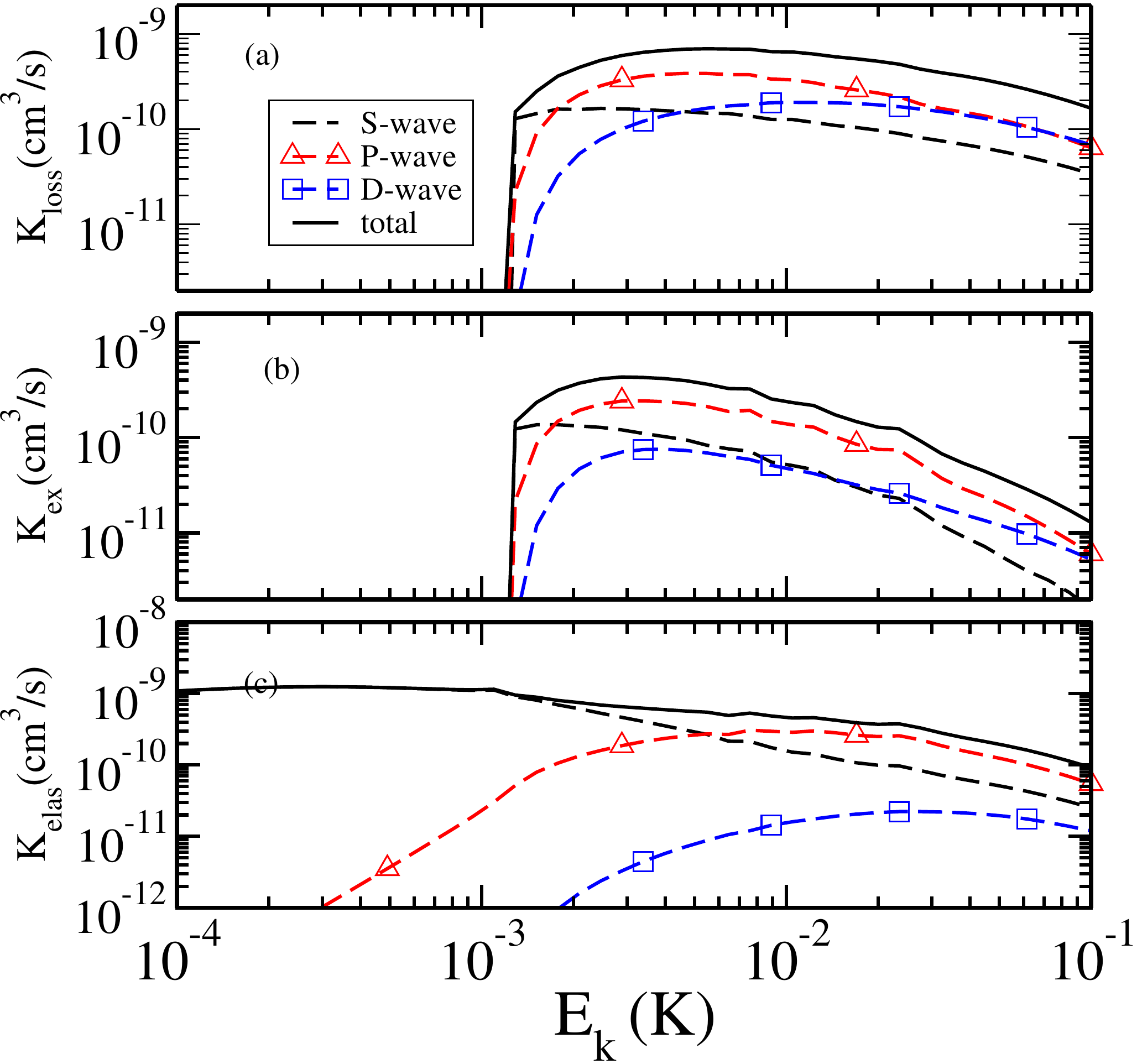}
\vspace{-.3cm}
\caption{
Reaction rates for $^{6}$Li + $^{4}$He$_2$, using zero-range (upper-set of panels) and 
finite-range (lower-set of panels) interactions, considering the (a2) model.}
\vspace{-0.5cm}
\label{fig05}
\end{figure}

For the model (a2), the $^{6}$Li + $^{4}$He$_2\to\, ^4$He + $^{4}$He$^{6}$Li reaction is
endothermic, as seen from the Table~\ref{tab3}. The corresponding results are given in Fig.~\ref{fig05}, where 
the atom exchange channel opens at 1.2mK and the dissociation at 1.3mK, offering an interesting interplay 
between the competing losses in these two channels. For the elastic reaction rate, the $s-$wave contribution is 
relevant below 1mK, with the $p-$wave being dominant above 10mK,  while the $d-$wave is marginal. 
In the atom exchange rate the $s-$ and $d-$waves are present, although the $p-$wave contributes to a large extend.

The loss rate  corresponds essentially to the atom exchange rate up to 5mK,  while above 10mK it receives 
contribution from the dissociation process in both $p-$ and $d-$waves, which attains values on the bold 
part of 10$^{-9}$cm$^3$/s. The relevant contributions from the $p-$ and $d-$waves amounts to the smallest 
of the binding energy of the $^4$He-$ ^6$Li molecule, which extends the Efimov long range potential, 
increasing the relevance of $p-$ and $d-$waves to the reaction process.

\section{Three body reactions with Helium-4 and Lithium-7}\label{sec5}
\subsection{ Reaction rates for $^{4}$He + $^4$He$^7$Li}
In this section, we report our results on the elastic, exchange and loss rates for the system containing the $^7$Li 
isotope, 
namely, when considering the $^{4}$He + $^4$He$^7$Li reaction, which has the diatomic and triatomic molecules  
more bound than the previous case where the Lithium-6 isotope was interacting with the other two $^4$He atoms.
For the inputs of the zero and finite-range models, the corresponding parameters for the potentials (a1) and (a2)
are given in Table \ref{tab1}.
For both sets (a1) and (a2)  the reaction $^{4}$He + $^4$He$^7$Li$\to ^4$He$_2$ + $ ^7$Li is endothermic, 
with the corresponding binding energies being quite close.  As verified,  
the results for the rates are very much similar, as it will be detailed in our discussion of Figs.~\ref{fig06} and \ref{fig07}. 
In this case, the separable potential has the scattering lengths much larger than the effective range, namely, about one order of magnitude and above, as seen in Table~\ref{tab2}, making the outcome of the zero-range model closer to the results 
obtained with the finite range potential, as the corrections due to the effective range are marginal.

In Fig.~\ref{fig06}, the reaction rates for the parameter set (a1) are shown, where the atom exchange channel opens around 
4.3mK, and close to this energy the $p-$wave elastic phase shift has a zero, clearly seen for the zero-range and finite 
range models. The 
$s-$wave amplitude gives the elastic rate up to the atom exchange threshold, when the $p-$wave becomes dominant, this is 
noticed by a depression in the rate, independent of the model.  The atom exchange rate is essentially defined by the $p-$wave, 
as we have discussed before. The loss rate in the $d-$wave becomes relevant at energies of about 100 mK.

\begin{figure}[htb]
\includegraphics[width=8cm]{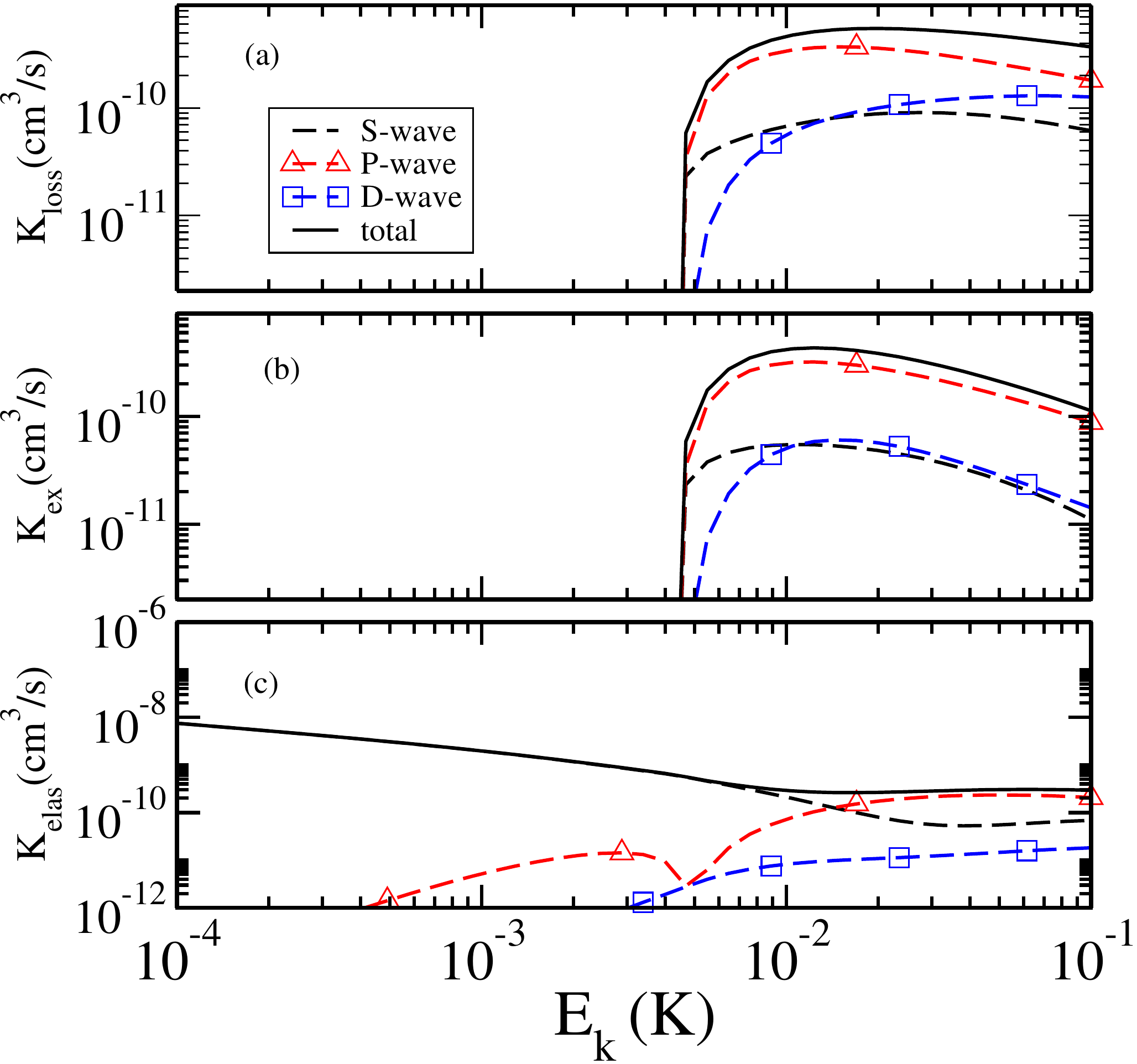}
\includegraphics[width=8cm]{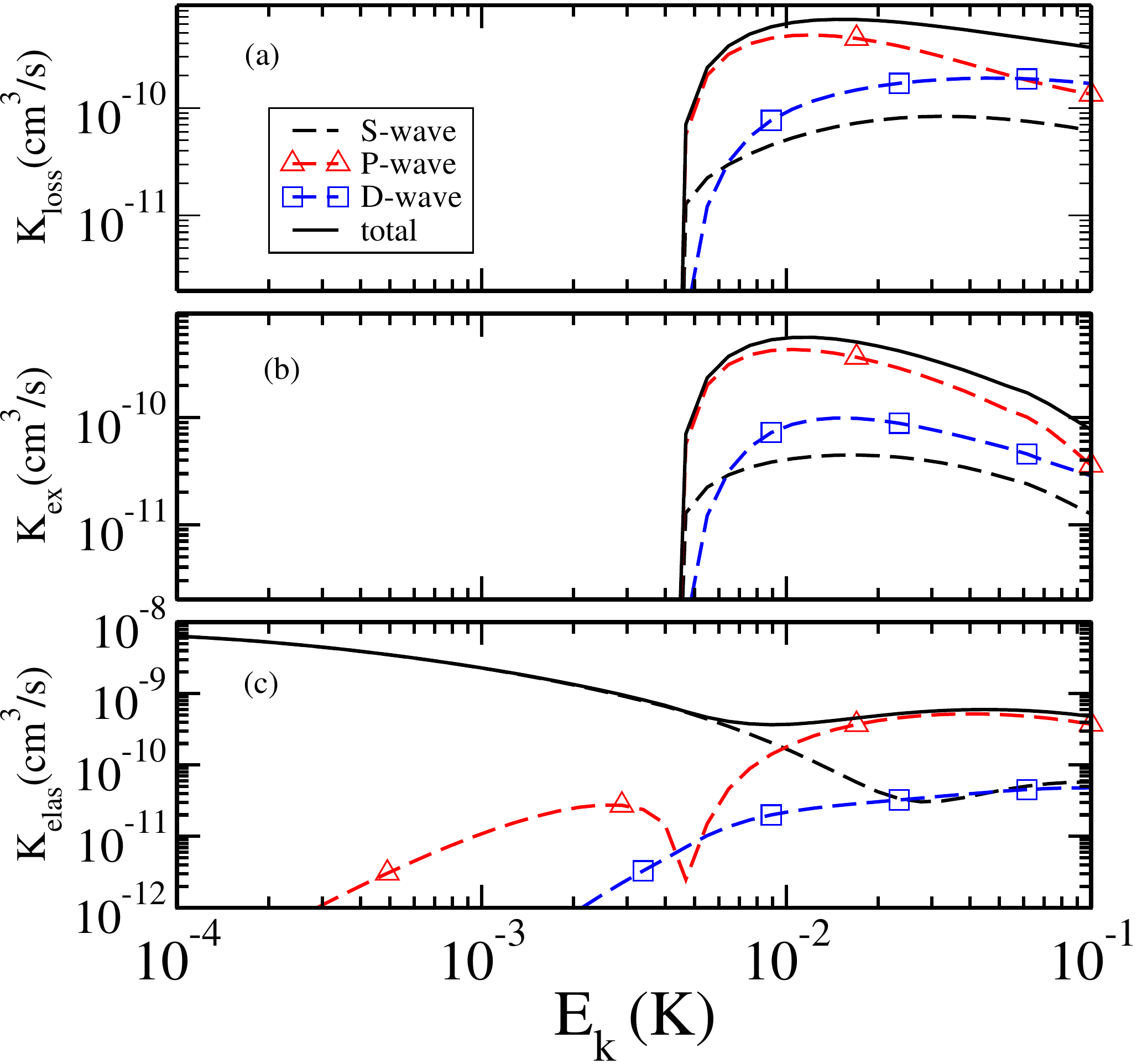}
\vspace{-.3cm}
\caption{
Reaction rates for $^{4}$He + $^4$He$^7$Li, using zero-range (upper-set of panels) and 
finite-range (lower-set of panels) interactions, considering the (a1) model.}
\vspace{-0.5cm}
\label{fig06}
\end{figure}

The calculations of the reaction rates with the parameter set (a2) are  shown in Fig. \ref{fig07} for the $^{4}$He + 
$^4$He$^7$Li collision. In this case the atom exchange reaction opens close to 1mK. Interesting enough is the zero of the 
elastic $p-$wave phase shift close to the threshold, as seen by the minimum of the elastic rate close to the atom exchange 
threshold. This minimum seems a universal property of the $p-$wave also appearing in the elastic process with model (a1). The 
difference between the two parameter sets is the binding energy of the  $^4$He$^7$Li molecule, as the triatomic binding 
energy is not relevant for the $p-$wave, which is also the protagonist on both the atom exchange and loss rate. The $s-$wave 
has a minor contribution on these two rates, and the $d-$wave appears in the loss rate above 100mK. The bulk of the results 
for the reaction rates are unaffected by the change of the potential range. 

\begin{figure}[h]
\includegraphics[width=8cm]{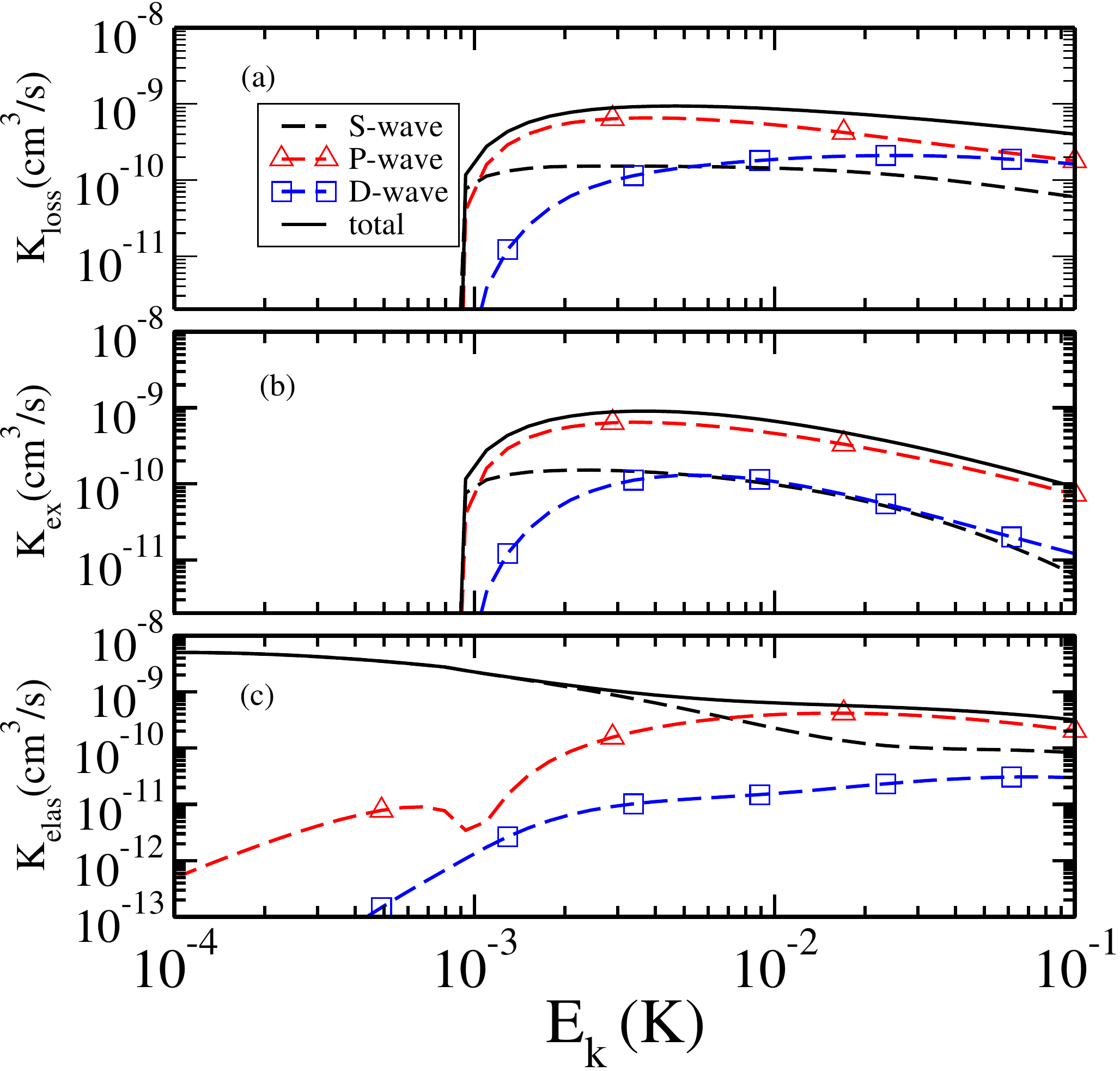}
\includegraphics[width=8cm]{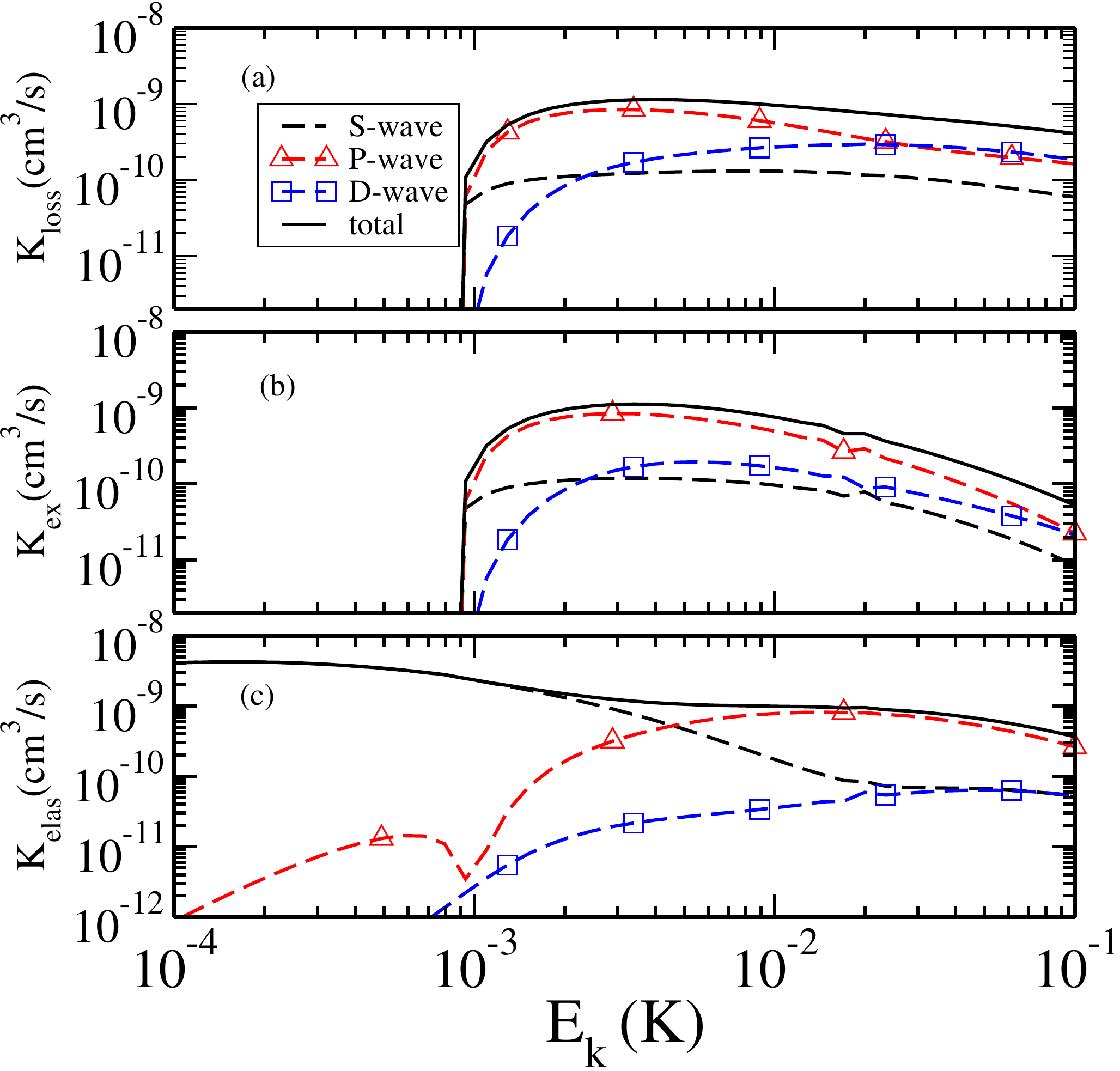}
\vspace{-.3cm}
\caption{
Reaction rates for $^{4}$He + ($^4$He$^7$Li), using zero-range (upper-set of panels) and 
finite-range (lower-set of panels) interactions, considering the (a2) model.}
\vspace{-0.5cm}
\label{fig07}
\end{figure}
\begin{figure}[h]
\includegraphics[width=8cm]{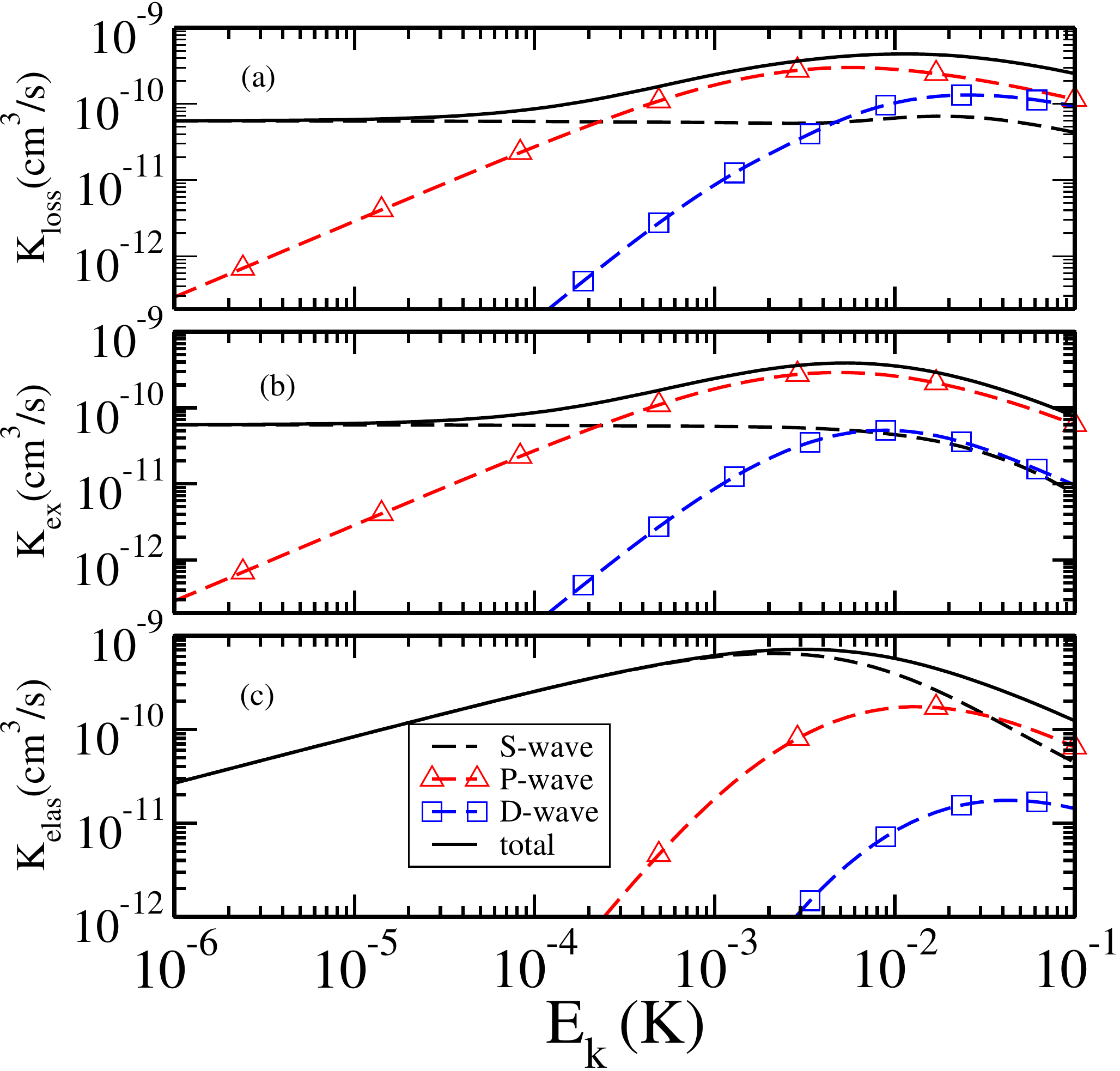}
\includegraphics[width=8cm]{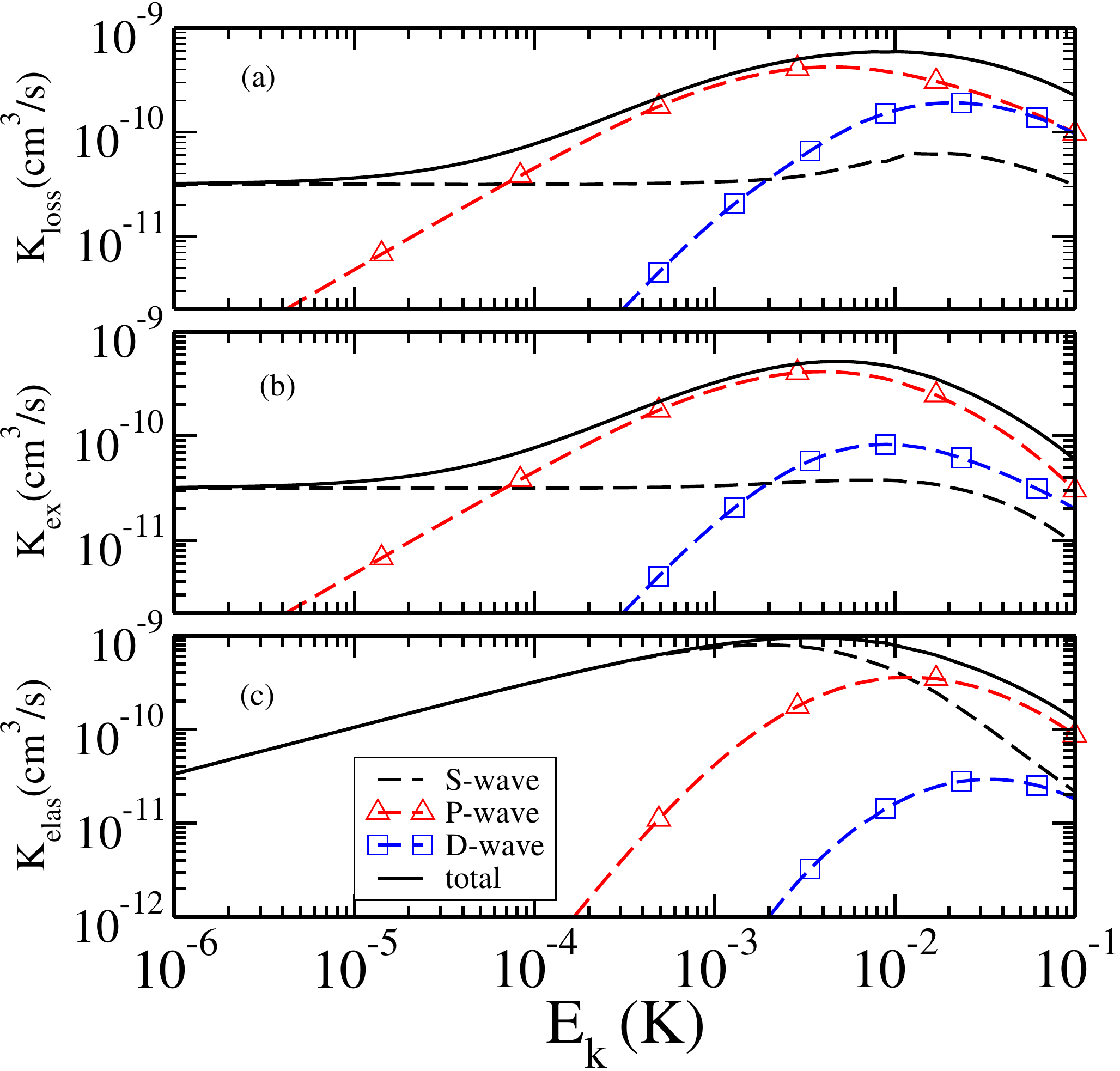}
\vspace{-.3cm}
\caption{
Reaction rates for $^{7}$Li + $^{4}$He$_2$, using zero-range (upper-set of panels) and 
finite-range (lower-set of panels) interactions, considering the (a1) model.}
\vspace{-0.5cm}
\label{fig08}
\end{figure}

\subsection{ Reaction rates for $^{7}$Li + $^{4}$He$_2$}\label{subsec7LiHe}
Our results on the reaction rates for $^{7}$Li + $^{4}$He$_2$, presented
in the Figs. \ref{fig08} and \ref{fig09}, are discussed in this subsection.  As represented in Table~\ref{tab2}, 
the atom exchange channel is exothermic for both set of potential models (a1) and (a2), with the 
dissociation threshold appearing at 1.3mK  for both two cases.

The results for the zero and finite-range models with parameters obtained from (a1), given in Table \ref{tab1}, are shown 
in Fig. \ref{fig08}. The $s-$wave dominates the elastic rate up to the dissociation threshold, when the $p-$wave takes over 
and gives the total value for the rate. The exothermic atom exchange reaction, i.e.,    $^{7}$Li + $^{4}$He$_2\to^4$He + 
$^4$He$^7$Li, has the major contribution from the  $s-$wave up to about 0.1mK, when the $p-$wave dominates. The loss rate 
shows the relevance of the $p-$wave up to 100mK, when the $d-$wave starts to compete.  Both zero-range and separable 
potential models provide very similar model independent results. So, for the results obtained in these calculations, it 
seems not relevant  the interaction range.

The next set of calculations were performed  for the zero and finite-range models with the parameters (a2), given in 
Table \ref{tab1}, and  presented in Fig. \ref{fig09}. Comparing with the binding energies from the set (a1), the corresponding 
values for the diatomic and triatomic molecules are more weakly bound, of about half of the corresponding values. This is 
reflected in the $s-$wave elastic rates that are about three times larger below 0.1mK, while the maximum is about the same and
somewhat below 10mK, with value of 10$^{-9}$cm$^3$/s. This is a consequence of the large difference in the binding 
energies of $^4$He$_ 2- ^7$Li molecule, which for set (a2) is considerably smaller than set (a1).  The $p-$wave outcome is 
only sensitive to the diatomic binding energies, with $^4$He$_ 2$ and $^4$He$ ^7$Li changing from 5.6 to 2.16 mK, that 
explains the results being quite close, as well as for the $d-$wave.

\begin{figure}[htb]
\includegraphics[width=8cm]{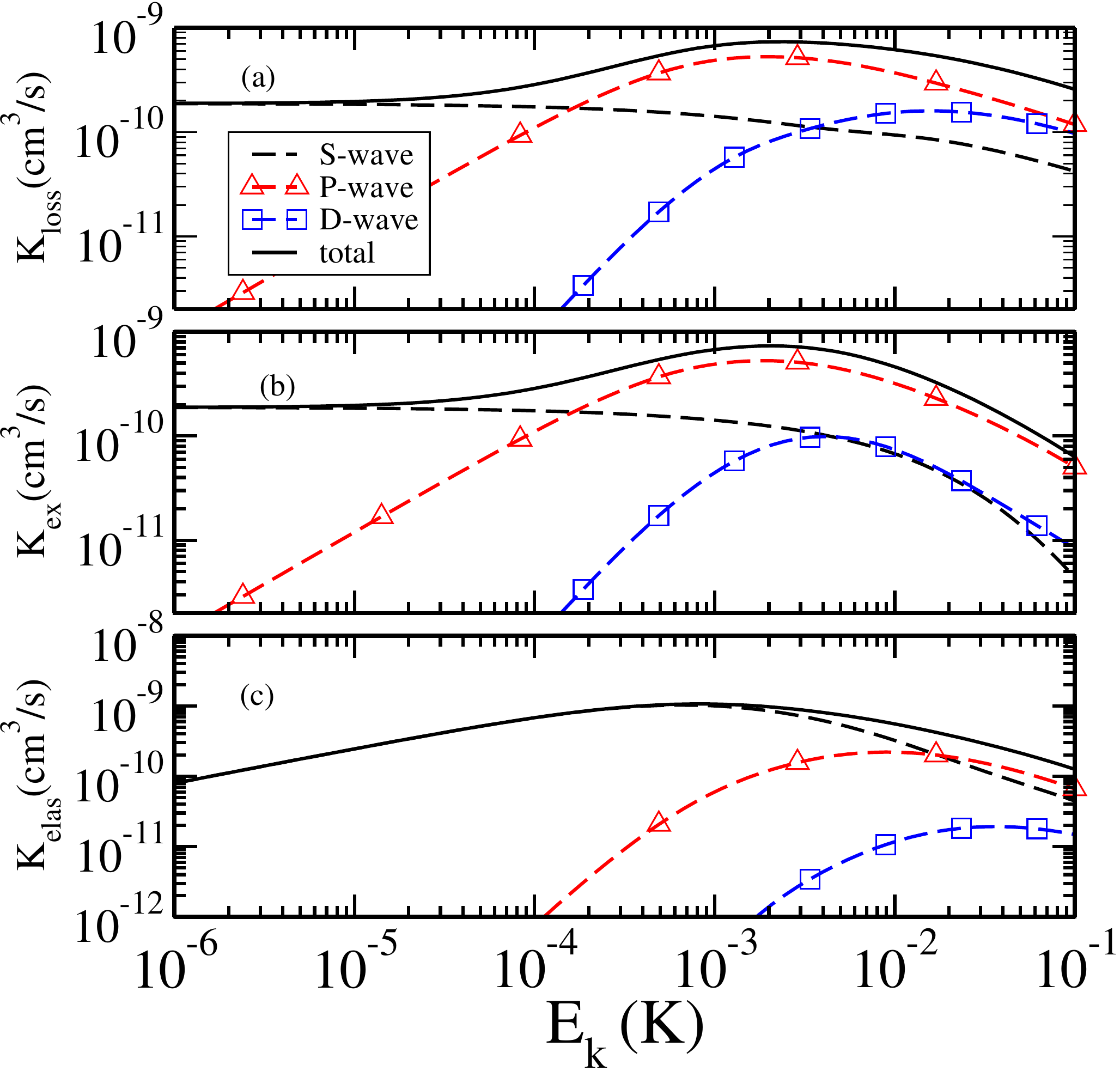}
\includegraphics[width=8cm]{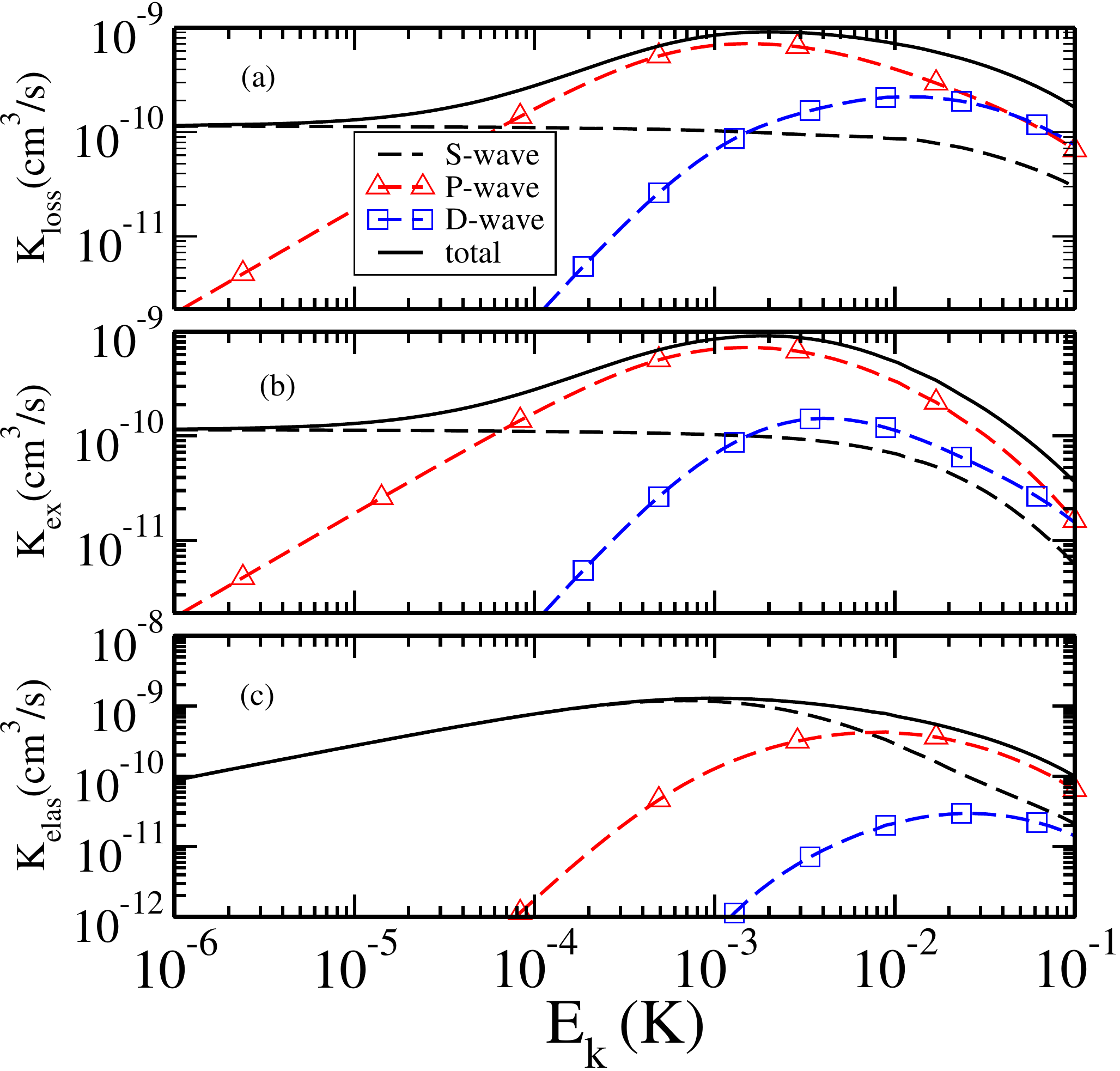}
\vspace{-.3cm}
\caption{
Reaction rates for $^{7}$Li + $^{4}$He$_2$, using zero-range (upper-set of panels) and 
finite-range (lower-set of panels) interactions, considering the (a2) model.}
\vspace{-0.5cm}
\label{fig09}
\end{figure}

The exothermic atom exchange reaction rate shown in Fig. \ref{fig09} has the dominance of the $s-$wave up to about 0.1mK, 
when the $p-$wave rises  and gives the bulk part of the rate. The $p-$wave also dominates the loss rate  up to 10mK, when the 
$d-$wave starts to compete.  The comparison between the zero-range and separable potential model results  shows again the 
model independence,  with the finite range  playing a minor role to build the bulk  values of the calculated rates.

\section{Three body reactions with Helium and Sodium}\label{sec6}
\subsection{Reactions rates for  $^{4}$He + $^4$He$^{23}$Na}
The collision $^{4}$He + $^4$He$^{23}$Na process has some distinctive features compared to the previous cases 
composed by helium and lithium atoms. Two salient differences will be apparent in the rate results. For this case, 
the parametrization of the separable potential gives an effective range, which is only about half of the scattering length,
as shown in Table~\ref{tab2}. The other feature is the presence of Efimov zeros in the elastic $s-$wave phase-shift~\cite{2018ShalchiPRA98}. 

\begin{figure}[htb]
\includegraphics[width=8cm]{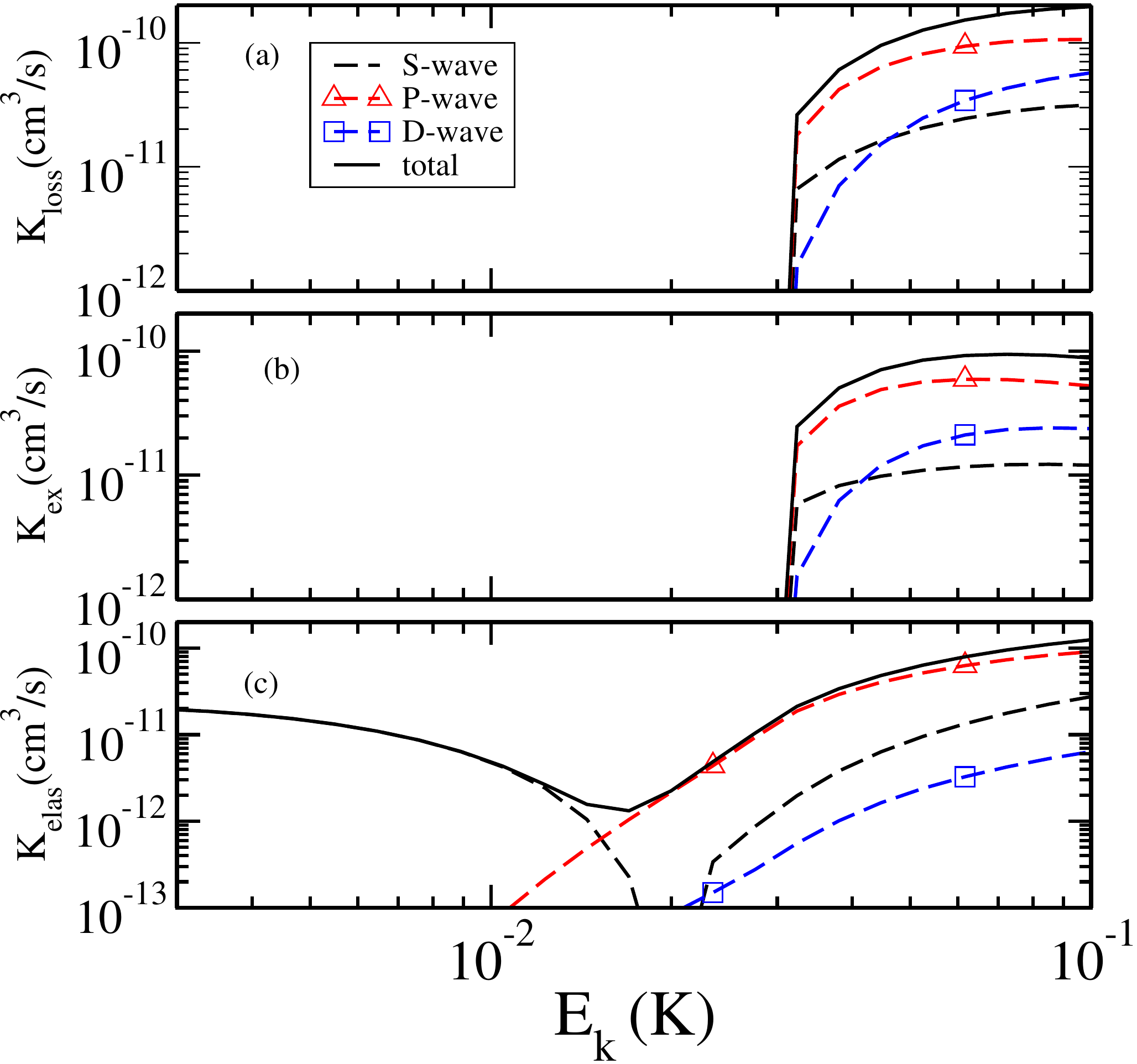}
\includegraphics[width=8cm]{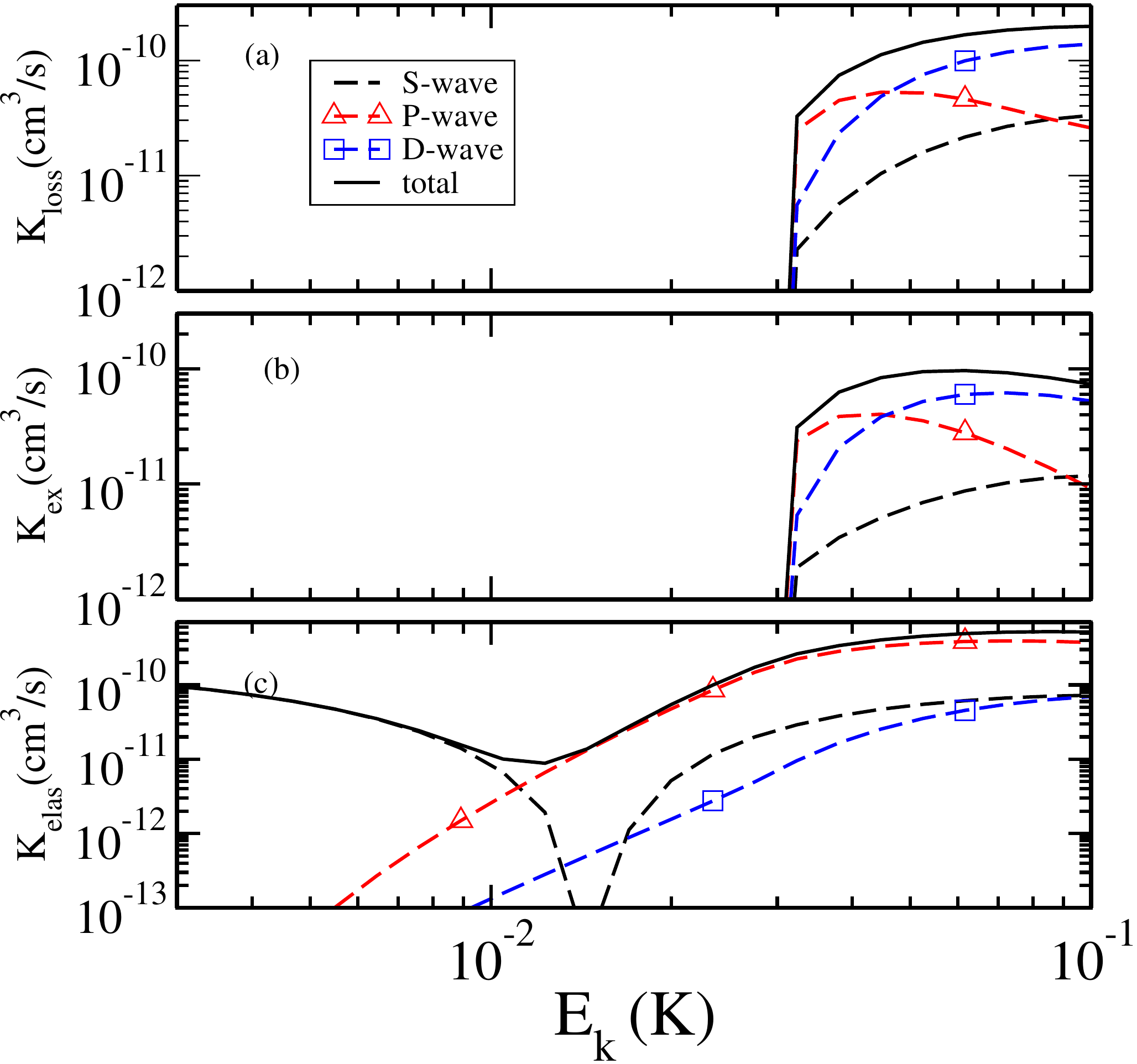}
\vspace{-.3cm}
\caption{
Reaction rates for $^{4}$He + $^4$He$^{23}$Na, using zero-range (upper-set of panels) and 
finite-range (lower-set of panels) interactions, considering the (a1) model.}
\vspace{-0.5cm}
\label{fig10}
\end{figure}

The Efimov zeros are a sequence of zeros/dips in the elastic scattering 
amplitude, that are controlled by both the scattering lengths and the actual value of the triatomic binding energy~\cite{2018ShalchiPRA98}.  The $^{4}$He + $^4$He$^{23}$Na $s-$wave phase-shift presents such a zero/dip, 
which is sensitive both to the two-atom low energy parameters and to the short range physics summarized in the 
binding energy of $^{4}$He$_2-^{23}$Na molecule. The zero turns into a dip, if it appears above the threshold of the 
atom-exchange or dissociation channels, due to the probability flux from the elastic channel to those ones.

\begin{figure}[htb]
\includegraphics[width=8cm]{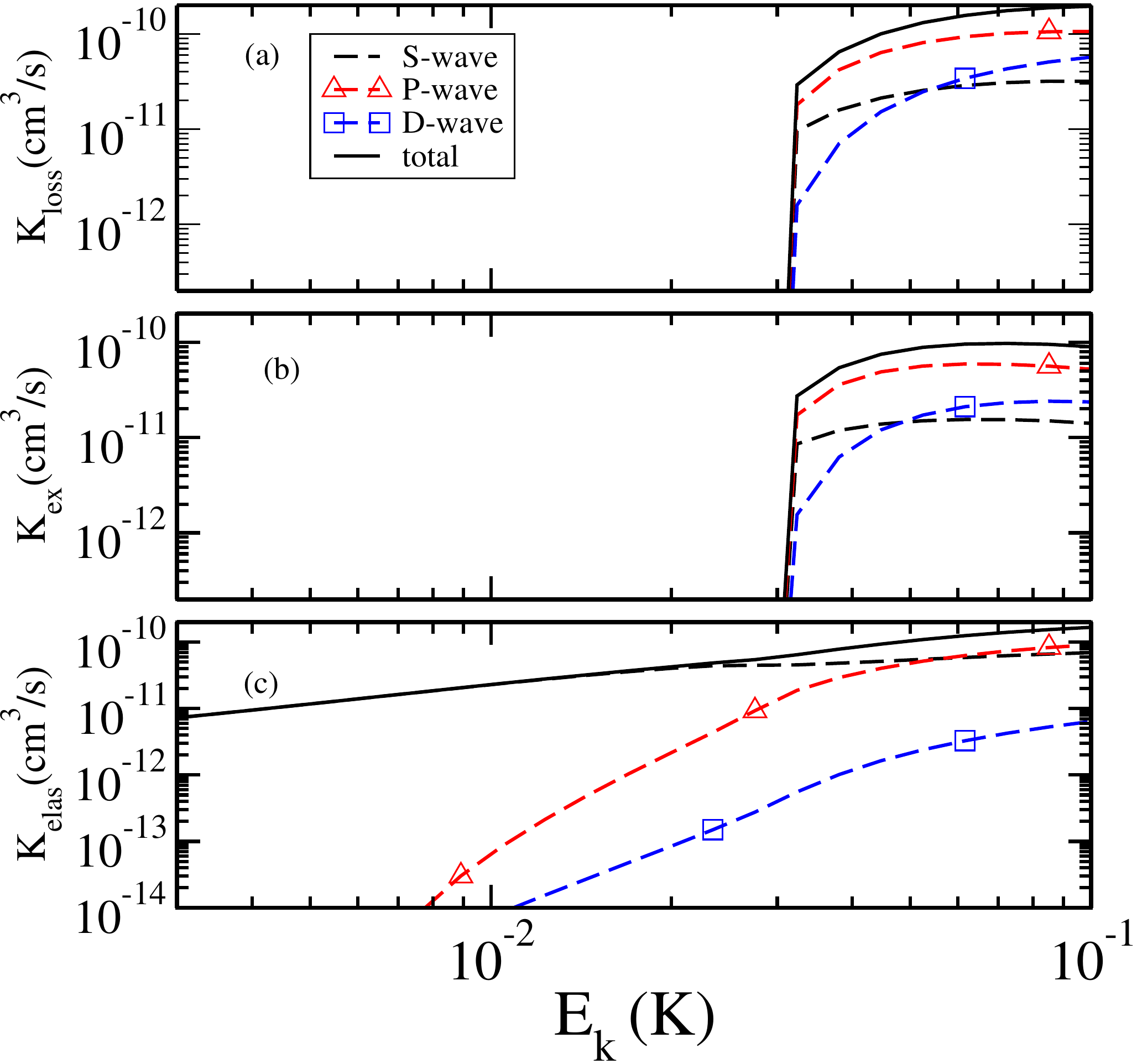}
\includegraphics[width=8cm]{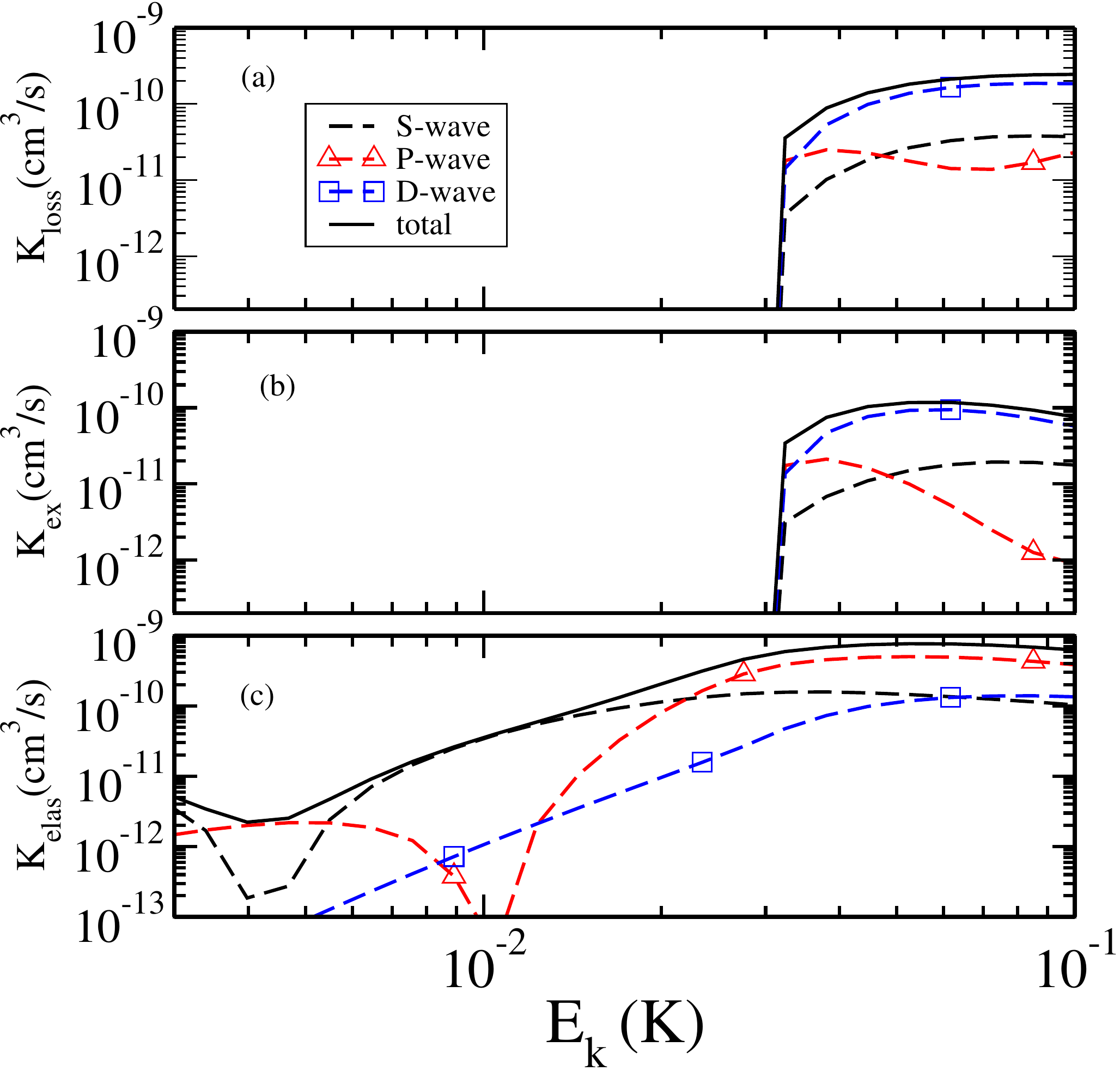}
\vspace{-.3cm}
\caption{
Reaction rates for $^{4}$He + $^4$He$^{23}$Na, using zero-range (upper-set of panels) and 
finite-range (lower-set of panels) interactions, considering the (a2) model.}
\vspace{-0.5cm}
\label{fig11}
\end{figure}

The atom-exchange reaction is now endothermic  for $^{4}$He + $^4$He$^{23}$Na$\to$  $^{4}$He$_{2}$ + $^{23}$Na 
and this channel opens around 28mK, while the dissociation reactions happens above  29mK,  for both  parameter sets (a1) 
and (a2) (see Table~\ref{tab3}). Our results for the reaction rates obtained with the zero-range and separable potential models  
with parameter sets (a1) and (a2)  are shown in Figs. \ref{fig10} and \ref{fig11}, respectively.

In Figs.~\ref{fig10} and  \ref{fig11}, the calculations of the reaction rates for the helium collision with ($^4$He$^{23}$Na) 
molecule are shown for both zero and finite range separable potentials. It is interesting to observe that the Efimov zero of the 
elastic $s-$wave phase-shift and corresponding reaction rate  appears for the zero-range and 
one-term Yamaguchi separable 
potentials respectively at 20mK and 15mK for the set (a1). In the case of the parameter set (a2), the zero of the elastic rate  for 
the potential comes around 2.5mK, while for the zero-range model the zero is not present in the scale of the figure.
The effective range moves the zero to somewhat larger values of the energy. These zeros are quite sensitive to the difference in 
the on-shell parameters as one can appreciate in Tables \ref{tab1} and \ref{tab2}.

The two potential models have identical diatomic and $^4$He$_2-^{23}$Na binding energies for the parameter set (a1), but 
they differ in the effective range and scattering lengths, which combined produce the change in the position of the $s-$wave 
zero, while keeping the same qualitative features in the elastic rate.  The zero of the $s-$wave elastic reaction rate gives together with 
the dominant $p-$wave contribution a pronounced dip, which for (a1) comes around 12-15mK, and the elastic rate from this 
energy up to 50 mK, increases two orders of magnitude  to about 10$^{-9}$cm$ ^3$/s. The same qualitative feature is found for 
the parameter set (a2), with the dip around 2.5mK for the separable 
potential, while for the zero-range model it is not present 
in the scale of the figure. The position of the minimum is sensitive to the effective range, as already observed for the parameter 
set (a1). For both parameter sets (a1) and (a2) the $p-$wave turns to be relevant between 10 and 30mK, while for the 
one-term separable potential with set (a2) it presents a zero.

The atom exchange and loss rates  are dominated by the higher waves, with particular preponderance of the $p-$wave, as found
 for the triatomic systems with the lithium isotopes. When considering  the sets (a1) and (a2) 
parametrized by Yamaguchi potentials, the influence of the effective range is verified in both cases 
by the importance of the $d-$wave with respect to the $p-$wave contribution. 
Naively, one can understand the effect of the effective range on the imbalance of the $p-$ and 
$d-$waves, as it cuts the long-range Efimov potential, which is relatively more important for the $p-$wave 
than for the $d-$wave. In the zero range model, without the weakening of the long range potential, the $p-$wave 
is dominant on the exchange and loss rates.
 
\subsection{Reaction rates for  $^{23}$Na + $^{4}$He$_2$}
The reaction $^{23}$Na + $^{4}$He$_2$ has an exothermic channel with the atom exchange to form the 
$^4$He$_2-^{23}$Na molecule (see Table~\ref{tab3}), and the dissociation threshold is quite low at 1.31mK, for both set 
of parameters (a1) and (a2).  The results are shown in Figs.~\ref{fig12} and \ref{fig13}, for both zero-range and one-term
separable Yamaguchi potentials with parameter sets (a1) and (a2), respectively. The magnitude of the reaction rates are 
found to grow up to 10$^{-10}$cm$^3$/s.  These values are one order  below the reaction rates obtained with the 
isotopes of lithium with mass numbers 6 and 7,  as we have throughly discussed in  
sections \ref{subsec6LiHe} and \ref{subsec7LiHe}.

The elastic reaction rate is dominated by the $s-$wave up to about 10mK, independent of the parameter set (a1) or (a2) and 
potential range, as shown in  Figs.~\ref{fig12} and \ref{fig13}. The $p-$wave importance above such energies increases over 
the $s-$ and $d-$waves. The atom exchange process is open for this entrance reaction channel, and the $p-$wave contributes 
to the bulk of this rate above  1mK, while the $d-$wave becomes more relevant with respect to the $p-$wave for the finite 
range potential. This  happens due to the fact that the finite range of the potential has the effect to deplete the 
$p-$wave rate, as we have discussed before.  This behavior is independent of the parameter set (a1) or (a2).

Finally the dissociation rate component of $K_\text{loss}$ is essentially determined by the $p-$ and $d-$waves, with the 
$s-$wave contributing marginally. The inversion of the importance of the $p-$ and $d-$waves to the dissociation process 
comes as a consequence of the consideration of the effective range. By analyzing separately the elastic and loss rates, 
one can disentangle the two and three-atomic low energy input informations contained $s-$, $p-$ and $d-$waves. Specifically,
the $s-$wave reaction rates depends on  two and three-atomic low energy input informations, while the higher partial waves
mainly carry the diatomic low energy parameters and being insensible to the triatomic molecular energy.
By   taking into account experimental differential angular reaction rates the individual
partial waves could be extracted.
 Following this procedure, real possibilities exist to separate the
effects of the triatomic molecule binding energy from the diatomic low energy informations, namely, binding energy, 
scattering length and effective range, which are contained in the elastic atom exchange and total loss rate from 
eventual experimental data below 100mK.

\begin{figure}[htb]
\includegraphics[width=8cm]{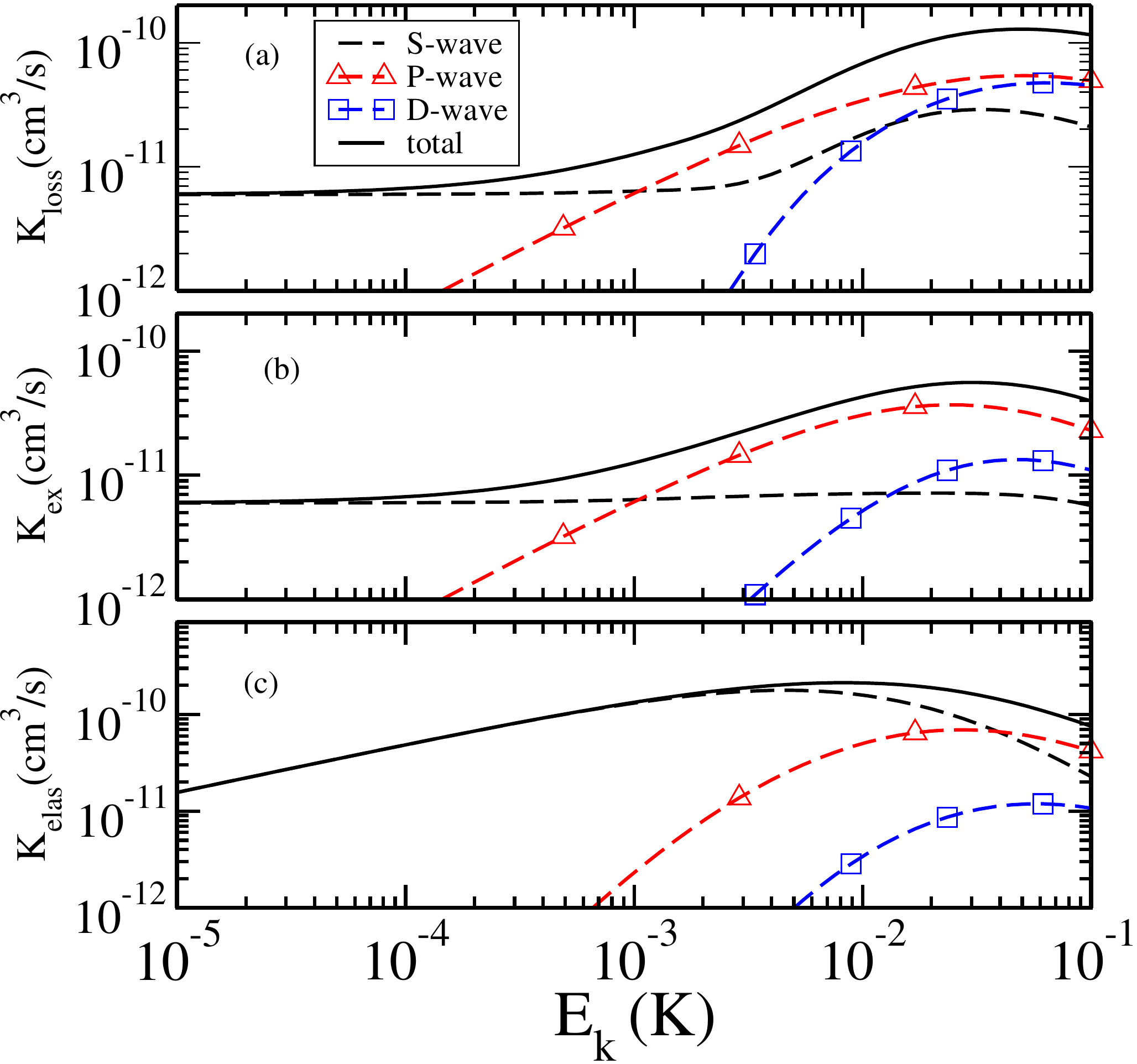}
\includegraphics[width=8cm]{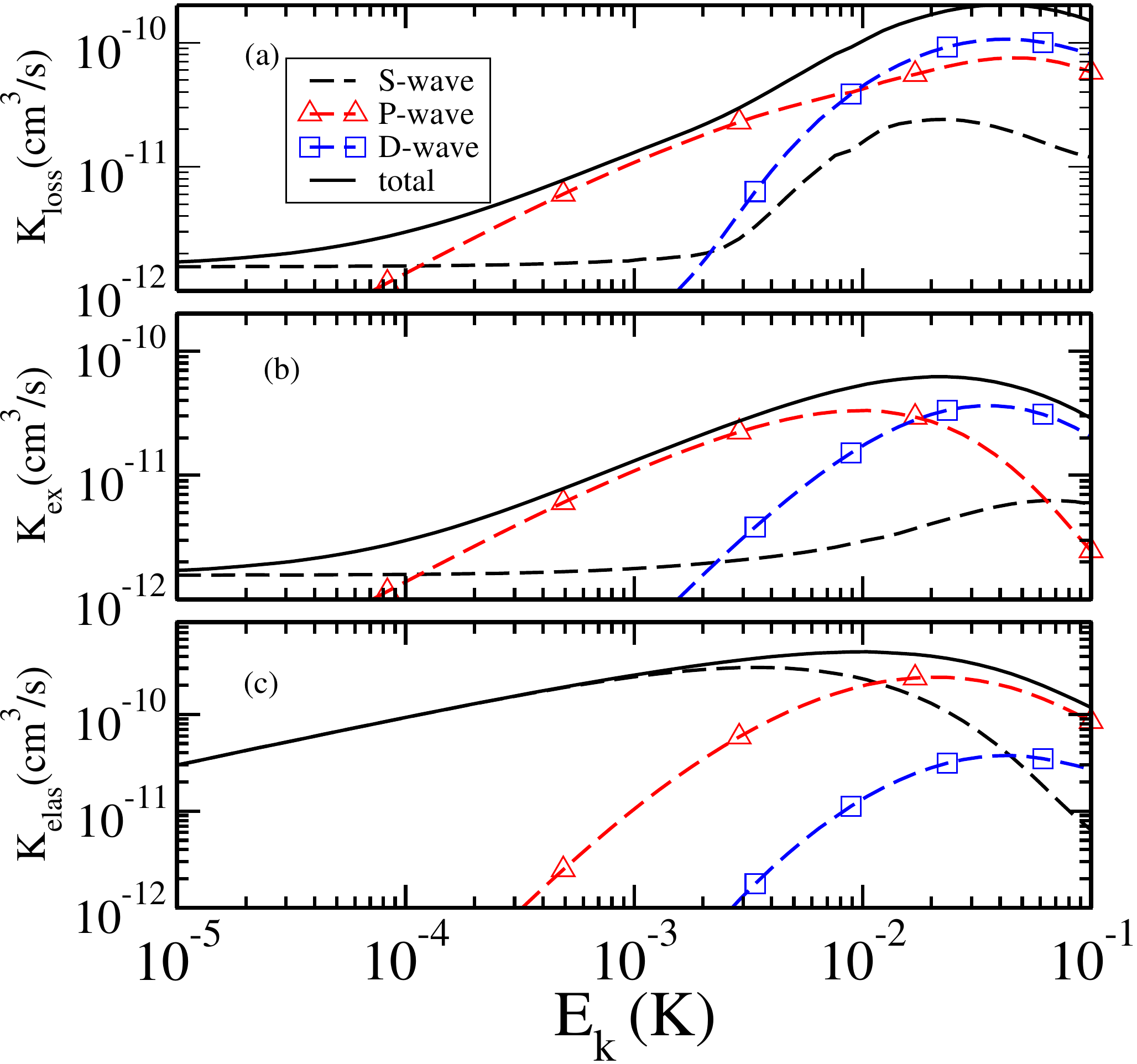}
\vspace{-.3cm}
\caption{
Reaction rates for $^{23}$Na + $^{4}$He$_2$, using zero-range (upper-set of panels) and 
finite-range (lower-set of panels) interactions, considering the (a1) model.}
\vspace{-0.6cm}
\label{fig12}
\end{figure}

\begin{figure}[htb]
\includegraphics[width=8cm]{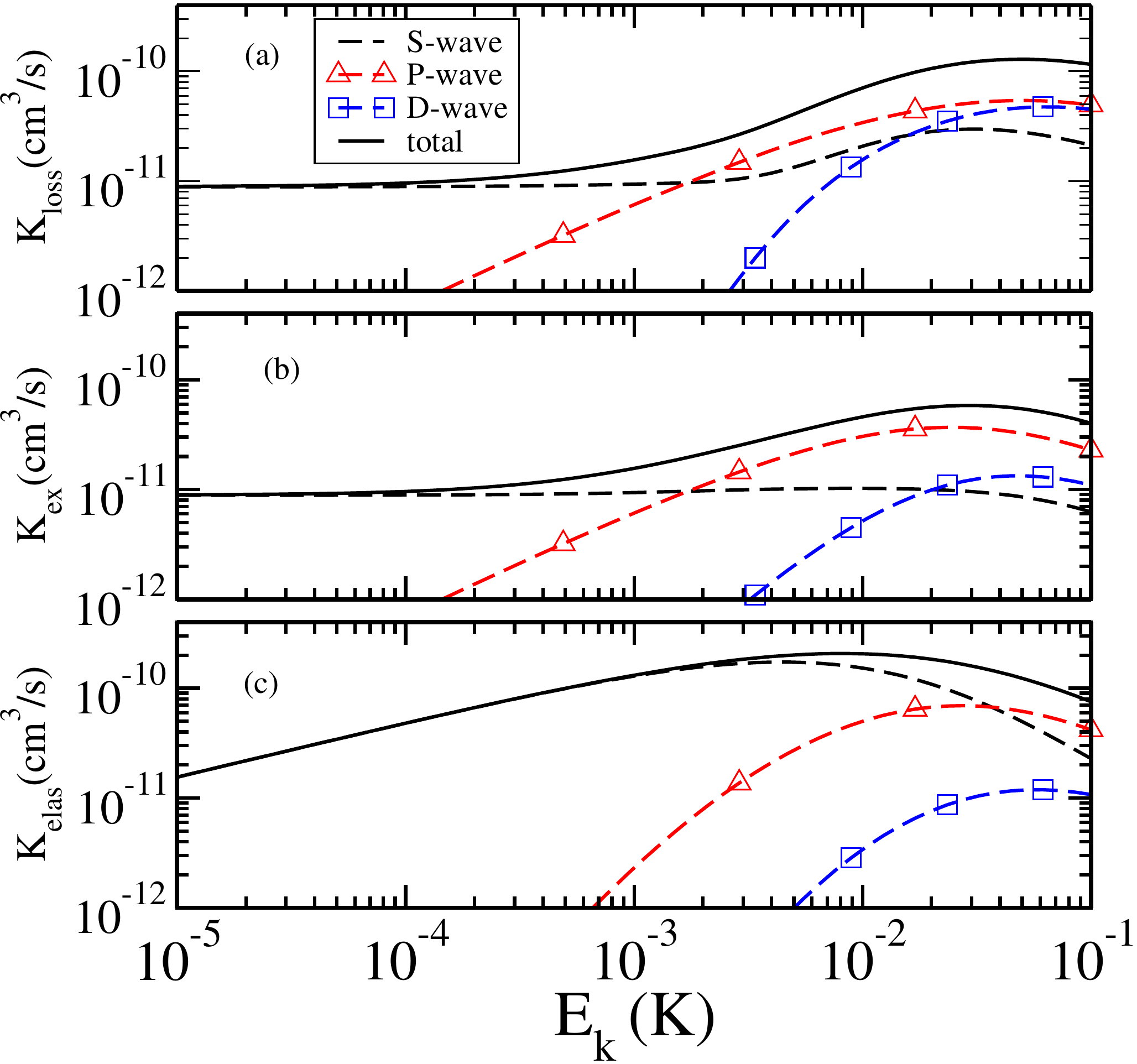}
\includegraphics[width=8cm]{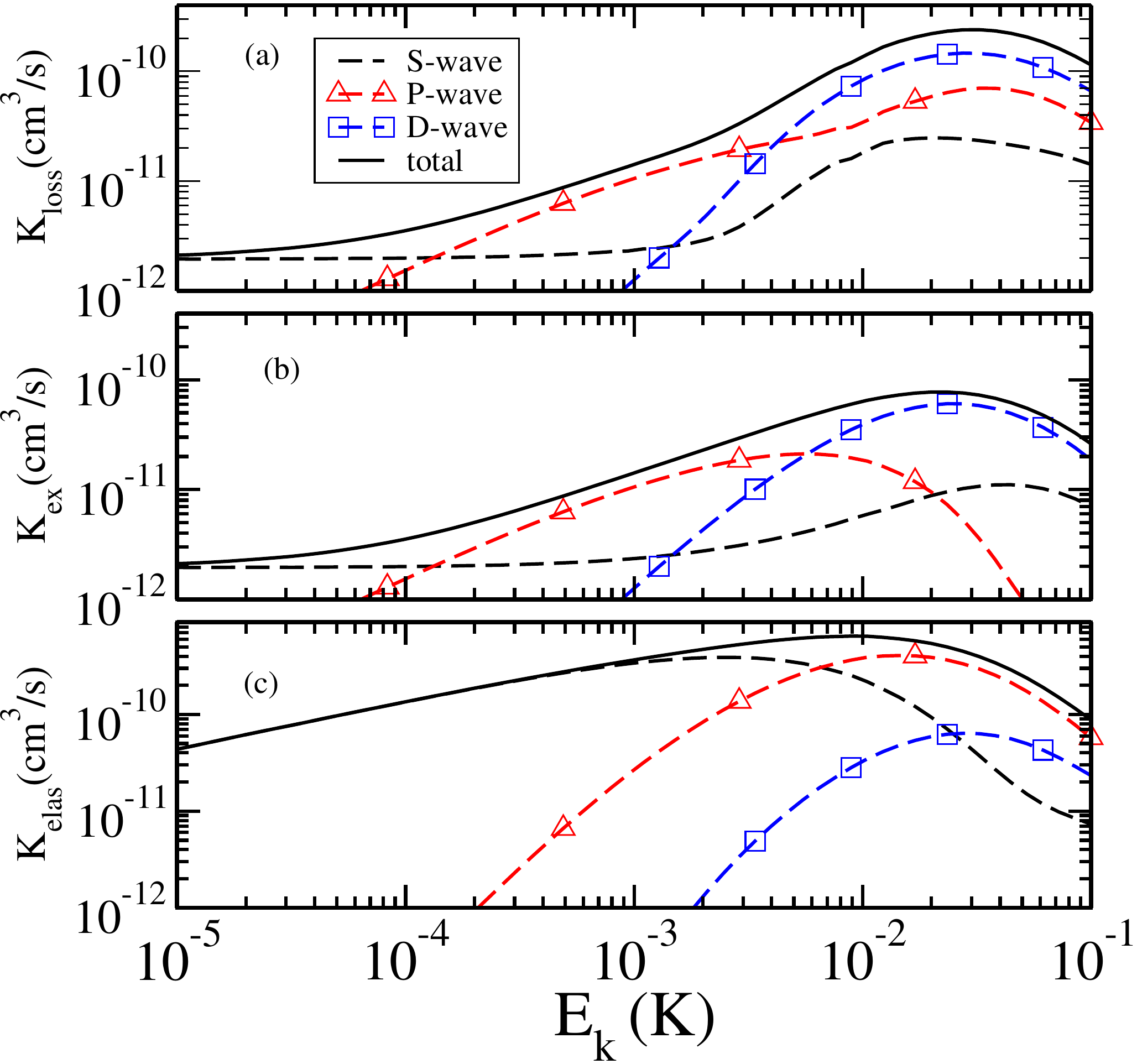}
\vspace{-.3cm}
\caption{
Reaction rates for $^{23}$Na + $^{4}$He$_2$, using zero-range (upper-set of panels) and 
finite-range (lower-set of panels) interactions, considering the (a2) model.}
\vspace{-0.5cm}
\label{fig13}
\end{figure}

\section {Conclusion} \label{conclusion}
In the present work, we provide predictions for several atom-dimer reaction rates, by considering particular two-atomic systems 
for different configurations having two $^4$He atoms with one of the species, among $^{6}$Li, $^{7}$Li and $^{23}$Na, 
taking the advantage that dimer and trimer binding energies are available for these atom-dimer systems from experimental
data or predicted by realistic potential model calculations.  Within the Faddeev formalism approach, these realistic binding 
energies are directly applied, 
in our study with zero-range and finite-range two-body interactions. We predict reaction rates for cold  $^4$He  elastic 
collisions with $^4$He$^{6}$Li, $^4$He$^{7}$Li and $^4$He$^{23}$Na molecules, as well as by considering the atomic 
species $^{6}$Li, $^{7}$Li and $^{23}$Na reacting with helium dimers for center-of-mass kinetic energies up to 100~mK. 

For our study we use the diatomic and triatomic  parameters from two atomic models given in \cite{2017Suno}. The elastic, 
atom exchange and loss rates are computed with a zero-range and a $s-$wave one-term separable potential in 
order to access the relevance of the effective range, besides the diatomic binding energies from \cite{2017Suno}. This work 
follows a previous exploration of the cold $s-$wave cross-sections for the $^4$He on the diatomic molecules formed by 
the helium atom with the isotopes  of lithium and sodium~\cite{2018ShalchiPRA98} using the zero-range and one-term 
separable potentials. 
 
Specifically, the elastic, atom-exchange and dissociation channels are investigated for the following six reactions:
 \begin{itemize}
\item $^{4}$He + $^4$He$^6$Li and  $^{6}$Li + $^{4}$He$_2$, in Section~\ref{sec4};
\item $^{4}$He + $^4$He$^7$Li and  $^{7}$Li + $^{4}$He$_2$, in Section~\ref{sec5};
\item $^{4}$He + $^4$He$^{23}$Na and $^{23}$Na + $^{4}$He$_2$, in Section~\ref{sec6}.
 \end{itemize}
The main characteristic found in these studies is the $p-$wave dominance of the atom-exchange and dissociation reaction, allowing 
 the separation of the effects of the on-shell low energy diatomic properties from the triatomic binding energy, which is
  determinant for the elastic reaction rate below the dissociation threshold.  In particular, for the $^{6,7}$Li and $^{23}$Na 
  reactions with the $^4$He$_2$ dimer, the atom exchange channel is always open, which allows to separate the $s-$wave 
  contribution below the dissociation threshold from the higher partial wave rates.
  
  Of possible experimental relevance,  is the presence of a minimum in the $s-$wave elastic reaction rate for the 
  $^4$He$\to(^4$He-$^{23}$Na) scattering, following  the previous findings of Ref.~\cite{2018ShalchiPRA98}. This minimum 
  is kept once the other partial waves are included in the elastic  rate, and thus could be of interest from the point of view 
  of an actual experiment. Such minimum is reminiscent of the Efimov effect that translates in the scattering region as a 
  log periodicity of zeros in the elastic $s-$wave phase shift in the collision of the atom with a weakly bound dimer.  
  Another interesting property is the presence of a zero/dip for some of the elastic rate in the $p-$wave, which was 
  particularly seen in the case of $^{4}$He + ($^4$He$^7$Li) and $^{4}$He + ($^4$He$^{23}$Na) between 1 and 10mK.
  
 The atom exchange and dissociation rates for all examples we have studied is dominated by the $p-$wave, as it is understood 
 from the importance of the one atom exchange mechanism, which gives more weight  to the process at backward angles. Pictorially, 
 one of the atoms in the initial molecule is picked-up by the incoming atom, and the newly formed molecule is propagating backwards.
 We remind that already the kernel of the integral equations and inhomogeneous term for the zero-range and $s-$wave one 
 term separable potential  corresponds to an atom exchange amplitude.

  The  range correction to the reaction rates  computed with the one-term separable potential, is particularly noticeable as an 
  inversion of the relative importance between the $p-$ and $d-$waves, with respect to the results obtained with the 
  zero-range model.  This effect comes from the softening of  the long-range Efimov type potential at the short distances 
  due to the finite range, still perceptible for the $p-$wave, while the $d-$wave is less sensitive to such modification 
  due to the stronger centrifugal barrier. Therefore, such an effect emphasizes even more the
  possibility to disentangle the on-shell low energy two-atom informations and the triatomic binding energy,  with the last one is
  determinant for the $s-$wave rates, but irrelevant for the higher wave contributions.
  The $p-$ and $d-$wave dominate the atom exchange 
  and dissociation rates,  while presenting different sensitivities to the low energy diatomic parameters.

Although the results for the reaction rates  present some sensitivity to the potential range 
when the fitted potential presents effective ranges comparable to the scattering length, which is found
particularly for the $^4$He-$^4$He$^{23}$Na, still the bulk values obtained with the zero range and 
the finite range potential are similar, supporting the robustness of our predictions.

Closing our summary,  we should point to some future directions of our investigation.  
The methods employed here can be extended to ultracold atoms in atomic traps to study 
controlled chemical reactions, by means of tuning the scattering lengths with Feshbach resonances, 
and also moving the triatomic state by induced few-body forces~\cite{2012-review,2020dePaula}. In addition,
more complex  reactions considering the collision of diatomic molecules widens the scope of our investigation.
Another perspective is the manipulation of the aspect ratio of the trap, changing the effective dimension in which the 
reaction takes place in a continuous way from three to two dimensions~\cite{2018aRosa,2018Abhishek,2019Garrido,2019Beane}, 
 therefore we hope that in the future not only the interaction can be tuned but also the effective dimension. 
 
\section*{Acknowledgments}
This work was partially supported by Funda\c {c}\~{a}o de Amparo \`{a} Pesquisa do Estado de S\~{a}o Paulo 
[2017/05660-0 (T.F. and L.T.),  2019/00153-8 (M.T.Y.)],  
Coordena\c c\~ao de Aperfei\c coamento de Pessoal de N\'\i vel Superior (M.A.S.),  
Conselho Nacional de Desenvolvimento Cient\'{\i}fico e Tecnol\'{o}gico [303579/2019-6 (M.T.Y.),
308486/2015-3 (T.F.), 304469-2019-0(L.T.), and Project INCT-FNA 464898/2014-5]. 

\appendix
\section{Atom-dimer transition operators}\label{app:atom-dimer-eqs}
As considering the particles $i$ and $j$ ($=1,2,3$), the general three-body scattering formalism, 
when the particle $i$ is the projectile, can be written in operator form  as~\cite{AGS,1983Glockle}
 \begin{eqnarray} \label{general}
  U_{ji}=\overline{\delta}_{ji}G_0^{-1}+\sum_{k}\overline{\delta}_{kj}t_{k}G_0 U_{ki}
 \end{eqnarray}
where $U_{ji}$s are transition amplitudes, $t_k$s are two body T-matrices and $G_0$ is three-body free propagator. 
If particle 1 is the projectile $U_{11}$ is related to elastic scattering amplitude and 
 $U_{21}$ and $U_{31}$ are related to rearrangement amplitudes. 
 By multiplying the both sides of Eq. \eqref{general} from right and left to $G_0$:
\begin{eqnarray}
  G_0 U_{11}G_0&=&G_0 t_{2} G_0 U_{21}G_0+G_0 t_{3} G_0 U_{31}G_0 \nonumber \\
   G_0 U_{21}G_0&=&G_0 +G_0 t_{1} G_0 U_{11}G_0+G_0 t_{3} G_0 U_{31}G_0 \nonumber \\
    G_0 U_{31}G_0&=&G_0+G_0t_{1} G_0 U_{11}G_0+G_0 t_{2} G_0 U_{21}G_0 
 \end{eqnarray}
If we use separable potential for two-body interactions, we will have the following two-body T-matrices.
\begin{eqnarray}
 t_{i}=\mid g_{i}\rangle \tau_{i}\langle g_{i}\mid \nonumber \\
 \langle  {\bf p\, q}\mid t_i\mid {\bf p'q'}\rangle&=&\delta({\bf p-p'})\delta({\bf q-q'})\tau(E_3-\frac{q^2}{2\mu_{iq}})\, ,
\end{eqnarray}
where $\mid g_{i}\rangle$ is called the form factor. By introducing the following operators:
\begin{eqnarray}
X_{ij}&=&\langle g_{i}\mid G_0 U_{ij} G_0\mid g_j\rangle ~~,\nonumber \\
Y_{ij}&=&\langle g_{i}\mid G_0\mid g_{j} \rangle 
\end{eqnarray}
we will have:
\begin{eqnarray}\label{general2}
  X_{11}&=&Y_{12}\tau_{2}X_{21}+Y_{13}\tau_{3}X_{31} \, ,\nonumber \\
  X_{21}&=&Y_{21}+Y_{21}\tau_{1}X_{11}+Y_{23}\tau_{3}X_{31} \, ,\nonumber \\
  X_{31}&=&Y_{31}+Y_{31}\tau_{1}X_{11}+Y_{32}\tau_{2}X_{21} \, .
 \end{eqnarray}
We consider $X_{ij}({\bf q,q'})=\langle {\bf q} \mid X_{ij}\mid {\bf q'}\rangle$ and 
 $Y_{ij}({\bf q,q'})=\langle {\bf q} \mid Y_{ij}\mid {\bf q'}\rangle$.
 For the scattering of a particle $\alpha$  by the $\alpha\beta$ bound subsystem, we have $1\rightarrow\alpha$, 
 $2\rightarrow\alpha$ and $3\rightarrow\beta$. Using symmetry property for the identical bosons, we consider 
 $ X_{11}({\bf q,q'})+X_{21}({\bf q,q'})=X({\bf q,q'})$ for elastic scattering and $X_{31}({\bf q,q'})=X'({\bf q,q'})$ for 
 rearrangement.  We can also calculate:
\begin{small}
\begin{eqnarray}
 Y_{13}({\bf q,q'})=Y_{31}({\bf q',q})=Y_{23}({\bf q,q'})=Y_{32}({\bf q',q})=\frac{K_1({\bf q,q'})}{2} \, ,\nonumber \\
 Y_{12}({\bf q,q'})=Y_{12}({\bf q',q})=Y_{21}({\bf q,q'})=Y_{21}({\bf q',q})=\frac{K_2({\bf q,q'})}{2}\, , \nonumber \\
\end{eqnarray}
\end{small}
 and we can write the partial wave decomposition of the coupled equations \eqref{general2}, as follows:
 \begin{eqnarray} \label{general3}
  X^\ell(q,q')&=&\frac{K^\ell_2(q,q')}{2}\nonumber \\
  &+&4\pi\int k^2 dk \frac{K^\ell_2(q,k)}{2}\tau_{\alpha\beta}(k;E_3)X^\ell(k,q') \nonumber \\
  &+&4\pi\int k^2 dk K^\ell_1(q,k)\tau_{\alpha\alpha}(k;E_3)X'^\ell(k,q') \, ,\nonumber \\
  X'^\ell(q,q')&=&\frac{K^\ell_1(q',q)}{2}\nonumber \\
  &&\hspace{-1cm} +4\pi\int k^2 dk \frac{K^\ell_1(k,q)}{2}\tau_{\alpha\alpha}(k;E_3)X^\ell(k,q')\, ,
 \end{eqnarray}
 where $\tau_1(E_3-\frac{k^2}{2\mu_{\alpha(\alpha\beta)}})=\tau_2(E_3-\frac{k^2}{2\mu_{\alpha(\alpha\beta)}})=
 \tau_{\alpha\beta}(k;E_3)$ and $\tau_3(E_3-\frac{k^2}{2\mu_{\beta(\alpha\alpha)}})=\tau_{\alpha\alpha}(k;E_3)$.
 By removing the singularity of two-body T-matrices:
 \begin{eqnarray}
  \tau_{\alpha\beta}(q;E_3)=\frac{\overline{\tau}_{\alpha\beta}(q;E_3)}{q^2-k_{\alpha}^2-{\rm i}\epsilon}\, ,\nonumber \\
  \tau_{\alpha\alpha}(q;E_3)=\frac{\overline{\tau}_{\alpha\alpha}(q;E_3)}{q^2-k_{\beta}^2-{\rm i}\epsilon}\, .
 \end{eqnarray}

We introduce the scattering and rearrangement reduced amplitudes as follow: 
 \begin{eqnarray}\label{def2}
  h^\ell_{\alpha}(q,q')=2\pi^2\overline{\tau}_{\alpha\beta}(q;E_3)X^\ell(q,q')\, ,\nonumber \\
 h^\ell_{\beta}(q,q')=4\pi^2\overline{\tau}_{\alpha\alpha}(q;E_3)X'^\ell(q,q')\, ,
 \end{eqnarray}
and using Eq. \eqref{def2} in Eq. \eqref{general3} we find the following  equations for the half-on-shell scattering 
and rearrangement reduced amplitudes:
 {\small
\begin{eqnarray}\label{eq:13}
&&h^\ell_{\alpha}(q;E_3)=\overline{\tau}_{\alpha}(q;E_3)\Bigg\{\frac{\pi}{2}K^\ell_2(q,k_\alpha;E_3)
+\int_0^{\infty} dk k^2\times \nonumber \\
&&\hspace{-.0cm}\times\Bigg[K^\ell_2(q,k;E_3) {h^\ell_{\alpha}(k;E_3)\over (k^2-k_\alpha^2-{\rm i}\epsilon)}
 +K^\ell_1(q,k;E_3)\frac{h^\ell_{\beta}(k;E_3)}{q^2-k_\beta^2-{\rm i}\epsilon}\Bigg]\Bigg\},\nonumber\\
&& h^\ell_{\beta}(q;E_3)=\overline{\tau}_{\beta}(q;E_3)\Bigg\{\frac{\pi}{2}K^\ell_1(k_\alpha,q;E_3)\nonumber\\
&&+\int_0^{\infty}dk k^2 K^\ell_1(k,q;E_3) {h^\ell_{\alpha}(k;E_3)\over(k^2-k_\alpha^2-{\rm i}\epsilon)}\Bigg\}\, ,
\end{eqnarray}
}
where $q'=k_{\alpha}=\sqrt{2\mu_{\alpha(\alpha\beta)}(E_3-E_{\alpha\beta})}$ and 
 \begin{eqnarray}\label{def1}
  \overline{\tau}_{\alpha}(q;E_3)=2\pi\overline{\tau}_{\alpha\beta}(q;E_3)\, ,\nonumber \\
 \overline{\tau}_{\beta}(q;E_3)=4\pi\overline{\tau}_{\alpha\alpha}(q;E_3)\, .
 \end{eqnarray}
 
  For the scattering of a particle $\beta$  by the $\alpha\alpha$ bound subsystem, we have $1\rightarrow\beta$, 
  $2\rightarrow\alpha$ and $3\rightarrow\alpha$. Using symmetry property for the identical bosons, 
 we consider $X'({\bf q,q'})\equiv X_{11}({\bf q,q'})$ for elastic scattering and $X({\bf q,q'})\equiv X_{21}({\bf q,q'})+
 X_{31}({\bf q,q'})$ for rearrangement.  We can also calculate:
\begin{small}
\begin{eqnarray}
 Y_{12}({\bf q,q'})=Y_{21}({\bf q',q})=Y_{13}({\bf q,q'})=Y_{31}({\bf q',q})=\frac{K_1({\bf q',q})}{2} \, ,\nonumber \\
 Y_{23}({\bf q,q'})=Y_{23}({\bf q',q})=Y_{32}({\bf q,q'})=Y_{32}({\bf q',q})=\frac{K_2({\bf q,q'})}{2}\, , \nonumber \\
\end{eqnarray}
\end{small}
 and we can write Eq.  \eqref{general2} as follows:
 \begin{eqnarray}\label{general4}
  X^\ell(q,q')&=&K^\ell_1(q,q')\nonumber \\
  &+&4\pi\int k^2 dk \frac{K^\ell_2(q,k)}{2}\tau_{\alpha\beta}(k;E_3)X^\ell(k,q') \nonumber \\
  &+&4\pi\int k^2 dk K^\ell_1(q,k)\tau_{\alpha\alpha}(k;E_3)X'^\ell(k,q') \nonumber \\
  X'^\ell(q,q')&=&4\pi\int k^2 dk \frac{K^\ell_1(k,q)}{2}\tau_{\alpha\alpha}(k;E_3)X^\ell(k,q') \nonumber \\
 \end{eqnarray}

Using Eq. \eqref{def2} in \eqref{general4} we have the following equations for the half-on-shell  scattering and 
rearrangement reduced amplitudes:
{\small
\begin{eqnarray}\label{eq:13p}
&&h^\ell_{\alpha}(q;E_3)=\overline{\tau}_{\alpha}(q;E_3)\Bigg\{\pi K^\ell_1(q,k_\alpha;E_3)
+\int_0^{\infty} dk k^2\times \nonumber \\
&&\hspace{-.0cm}\times\Bigg[K^\ell_2(q,k;E_3) {h^\ell_{\alpha}(k;E_3)\over (k^2-k_\alpha^2-{\rm i}\epsilon)}
 +K^\ell_1(q,k;E_3)\frac{h^\ell_{\beta}(k;E_3)}{q^2-k_\beta^2-{\rm i}\epsilon}\Bigg]\Bigg\},\nonumber\\
&& h^\ell_{\beta}(q;E_3)=\overline{\tau}_{\beta}(q;E_3)\int_0^{\infty}dk k^2 K^\ell_1(k,q;E_3) {h^\ell_{\alpha}(k;E_3)
\over(k^2-k_\alpha^2-{\rm i}\epsilon)}\, ,
\nonumber \\
\end{eqnarray}
}
here $q'=k_{\beta}=\sqrt{2\mu_{\beta(\alpha\alpha)}(E_3-E_{\alpha\alpha})}$, where $h_{\alpha}$ represents the 
rearrangement and $h_{\beta}$ represents the elastic scattering amplitudes.

\section{Scattering equation kernels}\label{app:kernels}
When using zero-range interactions, a momentum cut-off is required to regularize the integral equations for the 
$s-$wave state, within a  renormalization procedure, where the binding energy of the triatomic molecule is kept fixed. 
For that, in the kernels a subtraction is performed with a regularizing
momentum parameter $\mu$ (see e.g. \cite{2012-review}), such that the kernels $K_{1,2}$, $\overline{\tau}_{j}$ and 
$f_{\alpha,\beta}$ used in the bound-state and scattering equations
are given by 
{\small \begin{eqnarray}\label{eq:2p}
 &&K^\ell_{i=1,2}(q,k;E_3)\equiv G^\ell_i(q,k;E_3)-G^\ell_i(q,k,-\mu^2)\delta_{l0}, \nonumber\\
  &&G^\ell_1(q,k;E_3)=\int_{-1}^{1}dx\frac{P_\ell(x)}{E_3+{\rm i}\epsilon-\frac{q^2}{m}
- \frac{k^2}{2\mu_{\alpha\beta}}-\frac{kqx}{m}}
\nonumber\\
&&  G^\ell_2(q,k;E_3)=\int_{-1}^{1}dx\frac{P_\ell(x)}{E_3+{\rm i}\epsilon-\frac{q^2+k^2}{2\mu_{\alpha\beta}}
 -\frac{kqx}{Am}},
 \end{eqnarray}
 \begin{eqnarray}
 \overline{\tau}_{\alpha}(q;E_3)&\equiv& 
  \frac{\mu_{\alpha(\alpha\beta)}}{2\pi\mu_{\alpha\beta}^2}
 \left[\kappa_{\alpha\beta}+\kappa_{3,\alpha\beta}(E_3)
 \right],\\
 \overline{\tau}_{\beta}(q;E_3)&\equiv& 
 \frac{\mu_{\beta(\alpha\alpha)}}{2\pi\mu_{\alpha\alpha}^2}
 \left[\kappa_{\alpha\alpha}+\kappa_{3,\alpha\alpha}(E_3)
 \right] \, ,
 \end{eqnarray}}
where
\begin{eqnarray}
   \kappa_{\alpha\alpha}&\equiv&\sqrt{-2\mu_{\alpha\alpha} E_{\alpha\alpha}},
   \;\;\; \kappa_{\alpha\beta}\equiv\sqrt{-2\mu_{\alpha\beta}
   E_{\alpha\beta}}\nonumber \\
   \kappa_{3\alpha\alpha}(E_3)&\equiv&\sqrt{-2\mu_{\alpha\alpha}
   \left[E_3-   \frac{q^2}{2\mu_{\beta(\alpha\alpha)}}
      \right]},\nonumber \\
   \kappa_{3\alpha\beta}(E_3)&\equiv&\sqrt{-2\mu_{\alpha\beta}
   \left[E_3-\frac{q^2}{2\mu_{\alpha(\alpha\beta)}
   }\right]}.
\end{eqnarray}
\begin{eqnarray}
f_{\alpha}&=&\mu_{\alpha\beta}/\sqrt{k_{\alpha\beta}} \nonumber \\
f_{\beta}&=&\mu_{\alpha\alpha}/\sqrt{k_{\alpha\alpha}}
\end{eqnarray}

{\color{blue} In the case of the Yamaguchi separable potential from Eq.~\eqref{YamguchiV},}
$K_{1,2}$ and $\overline{\tau}_j$ are given by the following:
{\small \begin{eqnarray}
 K^\ell_1(q,k;E_3)&=&\int_{-1}^1 dx \left[q^2+\frac{k^2}{4}+qkx+\gamma_{\alpha\alpha}^2\right]^{-1}\\
 &\times&\left[k^2+\frac{q^2A^2}{(A+1)^2}+\frac{2qkAx}{(A+1)}+\gamma_{\alpha\beta}^2\right]^{-1}\nonumber \\
 &\times&\left[E_3+{\rm i}\epsilon-\frac{q^2}{m}-\frac{k^2}{2\mu_{\alpha\beta}}-\frac{qkx}{m}\right]^{-1}P_\ell(x),\nonumber
 \\
 K^\ell_2(q,k;E_3)&=&\int_{-1}^1\hspace{-0.2cm}dx\left[k^2+\frac{q^2}{(A+1)^2}+\frac{2qkx}{(A+1)}+
 \gamma_{\alpha\beta}^2\right]^{-1}\nonumber \\
 &\times& \left[q^2+\frac{k^2}{(A+1)^2}+\frac{2qkx}{(A+1)}+\gamma_{\alpha\beta}^2\right]^{-1}\\
 &\times&\left[E_3+{\rm i}\epsilon-\frac{(q^2+k^2)}{2\mu_{\alpha\beta}}-\frac{qkx}{Am}\right]^{-1}P_\ell(x),\nonumber
\end{eqnarray}}
\begin{eqnarray}
\overline{\tau}_{\alpha}(q;E_3)&\equiv& 
  \frac{\mu_{\alpha(\alpha\beta)}}{\pi\mu_{\alpha\beta}^2}\Bigg[
  \frac{\gamma_{\alpha\beta}(\gamma_{\alpha\beta}+\kappa_{\alpha\beta})^2}
{2\gamma_{\alpha\beta}+\kappa_{3\alpha\beta}(E_3)+\kappa_{\alpha\beta}}     \\
&\times&[\gamma_{\alpha\beta}+\kappa_{3\alpha\beta}(E_3)]^2
[\kappa_{\alpha\beta}+\kappa_{3\alpha\beta}(E_3)]\Bigg],
\nonumber \\
 \overline{\tau}_{\beta}(q;E_3)&\equiv&
   \frac{\mu_{\beta(\alpha\alpha)}}{\pi\mu_{\alpha\alpha}^2}\Bigg[\frac{
\gamma_{\alpha\alpha}(\gamma_{\alpha\alpha}+\kappa_{\alpha\alpha})^2
}
{2\gamma_{\alpha\alpha}+\kappa_{3\alpha\alpha}(E_3)+\kappa_{\alpha\alpha}}
\\
 &\times&
[\gamma_{\alpha\alpha}+\kappa_{3\alpha\alpha}(E_3)]^2 [\kappa_{\alpha\alpha}+\kappa_{3\alpha\alpha}(E_3)]\Bigg]
\nonumber.
\end{eqnarray}
\begin{eqnarray}
f_{\alpha}&=&2\pi\mu_{\alpha\beta}/\sqrt{k_{\alpha\beta}\gamma_{\alpha\beta}(k_{\alpha\beta}+
\gamma_{\alpha\beta})^3} \nonumber \\
f_{\beta}&=&2\pi\mu_{\alpha\alpha}/\sqrt{k_{\alpha\alpha}\gamma_{\alpha\alpha}(k_{\alpha\alpha}+
\gamma_{\alpha\alpha})^3}
\end{eqnarray}

In our approach, the  parameters of the separable interactions are fixed by the corresponding bound-state 
energies, as well as by the effective ranges (when considering finite-range interactions).

\section{Cross-sections}\label{app:xsection}
The scattering observables are obtained from the scattering amplitudes, which 
for elastic reaction is $U_{el}=U_{11}+U_{21}$ and for exchange reaction $U_{ex}=\sqrt{2}U_{31}$.
For the elastic scattering, we write that:
\begin{eqnarray}
 \frac{d\sigma_{el}}{d\Omega}=(2\pi)^4\mu_{\alpha(\alpha\beta)}^2|\langle \textbf{q}_f \phi_{\alpha}\mid U_{el}\mid 
 \textbf{q}_i\phi_{\alpha}\rangle|^2\, ,
\end{eqnarray}
where $\phi_{\alpha}$ is the bound state of $\alpha\beta$ subsystem and $q_i=q_f=
\sqrt{-2\mu_{\alpha(\alpha\beta)}(E_3-E_{\alpha\beta})}=\sqrt{-2\mu_{\alpha(\alpha\beta)}E_k}$.

We will use the following relations between two-body bound state and the form factor when we are using the 
one-term separable potential:
\begin{eqnarray}
V\mid\phi_{\alpha}\rangle=\lambda\mid g_{\alpha}\rangle\langle g_{\alpha}\mid \phi_{\alpha}\rangle=
f_{\alpha}\mid g_{\alpha}\rangle ;
\end{eqnarray}
consequently, in subsystem $\alpha\beta$ for on-shell momentum, we have:
\begin{eqnarray}
\tau_{\alpha\beta}(k;E_3)&=&
\frac{2\mu_{\alpha(\alpha\beta)} f_{\alpha}^2}{(k_{\alpha}^2-k^2)}~~~~,(k\rightarrow k_{\alpha})
\end{eqnarray}
and finally we can relate the transition amplitudes to our scattering function that we have calculate in Eq. \eqref{eq:13}
\begin{eqnarray}
 h_{\alpha}(\textbf{k};E_3)&=&2\pi^2\overline{\tau}_{\alpha\beta}(k;E_3)X(\textbf{k},\textbf{k}')\nonumber \\
 &=&-4\pi^2\mu_{\alpha(\alpha\beta)} f_{\alpha}^2X(\textbf{k},\textbf{k}')\nonumber \\
 &=&-4\pi^2\mu_{\alpha(\alpha\beta)} f_{\alpha}^2\langle \textbf{k}\mid\langle g_{\alpha} \mid G_0 U_{el} G_0\mid 
 g_{\alpha}\rangle \mid\textbf{k}'\rangle\nonumber \\
 &=&-4\pi^2\mu_{\alpha(\alpha\beta)} f_{\alpha}^2\langle \textbf{k} \phi_{\alpha} \mid V G_0 U_{el} G_0 V\mid 
 \textbf{k}' \phi_{\alpha} \rangle/ f_{\alpha}^2\nonumber \\
 &=&-4\pi^2\mu_{\alpha(\alpha\beta)} \langle \textbf{k} \phi_{\alpha} \mid U_{el}\mid \textbf{k}' \phi_{\alpha}  \rangle
\end{eqnarray}
 considering all the above equations the elastic cross section can be written as follow:
\begin{eqnarray}
\frac{d\sigma_{el}}{d\Omega}=|h_{\alpha}(\textbf{k};E_3)|^2 \, ,
\end{eqnarray}
For the exchange reaction cross section we have:
\begin{eqnarray}
 \frac{d\sigma_{ex}}{d\Omega}&=&(2\pi)^4\mu_{\beta(\alpha\alpha)}^2\sqrt{\frac{\mu_{\alpha(\alpha\beta)}}{\mu_{\beta(\alpha\alpha)}}}
 \sqrt{1-\frac{E_{\alpha\alpha}-E_{\alpha\beta}}{E_k}}
 \nonumber \\
 &\times &|\langle \textbf{q}_f \phi_{\beta}\mid U_{ex}\mid \textbf{q}_i\phi_{\alpha}\rangle|^2
\end{eqnarray}
where $q_f=\sqrt{2\mu_{\beta(\alpha\alpha)}(E_{\alpha\beta}-E_{\alpha\alpha}+E_k)}$, with the same method we can show that
\begin{eqnarray}
h_{\beta}(\textbf{k},E_3)=\frac{-8\pi^2\mu_{\beta(\alpha\alpha)}f_{\beta}}{\sqrt{2}f_{\alpha}}\langle 
\textbf{q}_f \phi_{\beta}\mid U_{ex}\mid \textbf{q}_i\phi_{\alpha}\rangle \, ,
\end{eqnarray}
resulting in:
\begin{equation}
 \frac{d\sigma_{ex}}{d\Omega}=\frac{f_{\alpha}^2}{2f_{\beta}^2}\sqrt{\frac{\mu_{\alpha(\alpha\beta)}}{\mu_{\beta(\alpha\alpha)}}}
 \sqrt{1-\frac{E_{\alpha\alpha}-E_{\alpha\beta}}{E_k}}|h_{\beta}(\textbf{k};E_3)|^2
  \, .
\end{equation}
In the case of scattering of $\beta$ from $\alpha\alpha$ we need scattering amplitudes
for elastic, $U_{el}=U_{11}$, and exchange channels, $U_{ex}=(U_{21}+U_{31})$, and the
cross-sections are written as:
\begin{eqnarray}
\frac{d\sigma_{el}}{d\Omega}=\frac{1}{4}|h_{\beta}(\textbf{k};E_3)|^2 \, ,
\end{eqnarray}
and 
\begin{equation}
 \frac{d\sigma_{ex}}{d\Omega}=\frac{f_{\beta}^2}{2f_{\alpha}^2}\sqrt{\frac{\mu_{\beta(\alpha\alpha)}}{\mu_{\alpha(\alpha\beta)}}}
 \sqrt{1-\frac{E_{\alpha\beta}-E_{\alpha\alpha}}{E_k}}|h_{\alpha}(\textbf{k};E_3)|^2 \, ,
\end{equation}
where $h_{\alpha}$ and $h_{\beta}$ are calculated from \eqref{eq:13p}.

\end{document}